\documentclass[structabstract]{aa}
\usepackage{natbib}        
\bibpunct{(}{)}{;}{a}{}{,} 
\usepackage{tabularx}
\usepackage{amsmath}       
\usepackage{multirow}

\usepackage{graphicx}
\usepackage{txfonts}

\newcommand {\nbodypp}{\textsc{\mbox{nbody6\raise.2ex\hbox{\tiny{++}}}}}

\newcommand {\Msun} {\mbox{M$_{\odot}$}}

\newcommand {\kms} {\mbox{km\,s$^{-1}$}}

\begin{document}
\title{Stellar interactions in dense and sparse star clusters}

\titlerunning{Stellar interactions in dense and sparse star clusters}

\author{C. Olczak\inst{1}
  \and S. Pfalzner\inst{1}
  \and A. Eckart\inst{1,2}}

\institute{I. Physikalisches Institut, Universit\"{a}t zu K\"{o}ln, Z\"{u}lpicher Str.77, 50937 K\"{o}ln, Germany \\
  \email{olczak@ph1.uni-koeln.de}
  \and Max-Planck-Institut f\"{u}r Radioastronomie, Auf dem H\"{u}gel 69, 53121 Bonn, Germany}

\date{Received ; accepted}

%
%

\abstract
{Stellar encounters potentially affect the evolution of the protoplanetary discs in the Orion Nebula Cluster (ONC). However, the role of encounters
  in other cluster environments is less known.}
{We investigate the effect of the encounter-induced disc-mass loss in different cluster environments.}
{Starting from an ONC-like cluster we vary the cluster size and density to determine the correlation of collision time scale and disc-mass loss. We
  use the {\textsc{\mbox{nbody6\raise.2ex\hbox{\tiny{++}}}}} code to model the dynamics of these clusters and analyze the disc-mass loss due to
  encounters.}
{We find that the encounter rate depends strongly on the cluster density but remains rather unaffected by the size of the stellar population. This
  dependency translates directly into the effect on the encounter-induced disc-mass loss. The essential outcome of the simulations are: i) Even in
  clusters four times sparser than the ONC the effect of encounters is still apparent. ii) The density of the ONC itself marks a threshold: in less
  dense and less massive clusters it is the massive stars that dominate the encounter-induced disc-mass loss whereas in denser and more massive
  clusters the low-mass stars play the major role for the disc mass removal.}
{It seems that in the central regions of young dense star clusters -- the common sites of star formation -- stellar encounters do affect the
  evolution of the protoplanetary discs. With higher cluster density low-mass stars become more heavily involved in this process. This finding
  allows for the extrapolation towards extreme stellar systems: in case of the Arches cluster one would expect stellar encounters to destroy the
  discs of most of the low- and high-mass stars in several hundred thousand years, whereas intermediate mass stars are able to retain to some extant
  their discs even under these harsh environmental conditions.}

\keywords{stellar dynamics -
  methods: N-body simulations -
  stars: pre-main sequence, circumstellar matter}

\maketitle

%

\section{Introduction}

To current knowledge, planetary systems form from the accretion discs around young stars. These young stars are in most cases not isolated but are
part of a cluster \citep[e.g.][]{2003ARA&A..41...57L}. Densities in these cluster environments vary considerably, spanning a range of 10\,pc$^{-3}$
(e.g. $\eta$ Chameleontis) to $10^6$\,pc$^{-3}$ (e.g. Arches Cluster). Though it is known that discs disperse over time
\citep{2001ApJ...553L.153H,2002astro.ph.10520H,2006ApJ...638..897S,2008ApJ...672..558C} and that in dense clusters ($n$\,$\gtrsim$\,$10^3$\,pc$^{-3}$)
the disc frequency seems to be lower in the core \citep[e.g.][]{2007ApJ...660.1532B}, it is an open question as to how far interactions with the
surrounding stars influence planet formation in clusters of different densities.

Two major mechanisms are potentially able to strongly affect the evolution of protoplanetary discs and planets in a cluster environment:
photoevaporation and gravitational interactions. Photoevaporation causes the heating and evaporation of disc matter by the intense UV radiation from
massive stars. First models of photoevaporation were developed by \citet{1998ApJ...499..758J} and \citet{1999ApJ...515..669S} (see also references in
Hollenbach, Yorke \& Johnstone 2000), and have been much improved in the past years
\citep{2001MNRAS.328..485C,2003ApJ...582..893M,2004RMxAC..22...38J,2005MNRAS.358..283A,2006MNRAS.369..229A,2008ApJ...688..398E,2009ApJ...690.1539G,2009ApJ...699L..35D}.
Gravitational interactions are another important effect on the population of stars, discs, and planets in a cluster environment. An encounter between
a circumstellar disc and a nearby passing star can lead to significant loss of mass and angular momentum from the disc. While such isolated encounters
have been studied in a large variety
\citep{1993ApJ...408..337H,1993MNRAS.261..190C,1994ApJ...424..292O,1995ApJ...455..252H,1996MNRAS.278..303H,1997MNRAS.287..148H,2004ApJ...602..356P,2005ApJ...629..526P,2006ApJ...653..437M,2007ApJ...656..275M,2008A&A...487..671K},
only few numerical studies have investigated the effect of stellar encounters on circumstellar discs in a dense cluster environment directly
\citep{2001MNRAS.325..449S,2006ApJ...641..504A}.

Only recently has it been shown from numerical simulations that stellar encounters do have an effect on the discs surrounding stars in a young dense
cluster \citetext{\citealp[][]{2006ApJ...642.1140O,2006A&A...454..811P,2006ApJ...652L.129P,2006ApJ...653..437M,
    2007ApJ...661L.183M,2007ApJ...656..275M,2007A&A...462..193P,2007A&A...475..875P}; \citealp[see also the review by][]{2007ARA&A..45..481Z}}. The
massive stars in the centre of such a stellar cluster act as gravitational foci for the lower mass stars \citep{2006A&A...454..811P}. Discs are most
affected when the masses of the stars involved in an encounter are unequal \citep{2006ApJ...642.1140O,2007ApJ...656..275M}, so it is the massive stars
that dominate the encounter-induced disc-mass loss in young dense clusters \citep{2006ApJ...642.1140O}. Observational evidence for this effect has
been found by \cite{2008A&A...488..191O}.

The numerical results obtained in our previous investigations are based on a dynamical model of the Orion Nebula Cluster (ONC) -- one of the
observationally most intensively studied young star cluster. It was demonstrated that in the ONC stellar encounters can have a significant impact on
the evolution of the young stars and their surrounding discs
\citep{2006ApJ...652L.129P,2006ApJ...642.1140O,2006A&A...454..811P,2007A&A...462..193P,2008A&A...488..191O,2008A&A...487L..45P}. However,
investigating one model star cluster is not sufficient to draw general conclusions -- in fact, one could not answer questions as: How would things
change in a \emph{denser} cluster? Would a higher density inevitably imply that stellar encounters play a more important role in the star and planet
formation process? And what would be the situation in more \emph{massive} clusters? Would the larger number of stars -- in particular \emph{massive}
stars -- play a role?

Conclusive answers to these questions demand further numerical investigations covering a larger parameter space in cluster parameters. Here we realise
this by modelling scaled versions of the standard ONC model -- clusters with varying stellar numbers and sizes. The focus of this investigation is on
the encounter-induced disc-mass loss. Throughout this work we assume that initially all stars are surrounded by protoplanetary discs. This is
justified by observations that reveal disc fractions of nearly 100\,\% in very young star clusters
\citep[e.g.][]{2000AJ....120.1396H,2000AJ....120.3162L,2001ApJ...553L.153H,2005astro.ph.11083H}. In Section~\ref{sec:observations_onc} we briefly
outline the observationally determined basic properties of the ONC, that serves as a reference model for the other cluster models. The computational
method is described in Section~\ref{sec:numerical_method} and the construction of the numerical models is outlined in
Section~\ref{sec:numerical_models}. Afterwards we present results from a numerical approach to this problem in Section~\ref{sec:numerical_results} and
compare with analytical estimates in Section~\ref{sec:analytical_results}. The conclusion and discussion mark the last section of this paper.

%

\section{Structure and dynamics of the ONC}

\label{sec:observations_onc}

The ONC is a rich stellar cluster with about 4000 members with masses $m$\,$\ge$\,0.08\,{\mbox{M$_{\odot}$}} and a radial extension of $\sim$2.5\,pc
\citep{1998ApJ...492..540H,2000ApJ...540..236H}. Most of the objects are T~Tauri stars. The mean stellar mass is about
$\bar{m}$\,$\approx$\,0.5\,{\mbox{M$_{\odot}$}} and the half-mass radius $R_{\rm{hm}}$\,$\approx$\,1\,pc \citep{1998ApJ...492..540H}. Recent studies
of the stellar mass distribution \citep{2000ApJ...540..236H,2000ApJ...540.1016L,2002ApJ...573..366M,2004ApJ...610.1045S} reveal no significant
deviation from the generalized IMF of \citet{2002Sci...295...82K},

\begin{equation}
  \xi(m)=
  \begin{cases}
    \, m^{-1.3} & , \quad 0.08 \le m/{\mbox{M$_{\odot}$}} <0.50 \,, \\
    \, m^{-2.3} & , \quad 0.50 \le m/{\mbox{M$_{\odot}$}} <1.00 \,, \\
    \, m^{-2.3} & , \quad 1.00 \le m/{\mbox{M$_{\odot}$}} <\infty \,.
  \end{cases}
  \label{eq:kroupa_imf}
\end{equation}

The mean age of the whole cluster has been estimated to be $t_{\rm{ONC}}$\,$\approx$\,1\,Myr, although a significant age spread of the individual
stars is evident \citep{1997AJ....113.1733H,2000ApJ...540..255P}.

The density and velocity distribution of the ONC resembles an isothermal sphere. The central number density $\rho_{\rm{core}}$ in the inner 0.053\,pc
reaches $4.7$\,$\cdot$\,$10^4$\,pc$^{-3}$ \citep{2002Msngr.109...28M} and makes the ONC the densest nearby ($<$\,1\,kpc) young stellar cluster. The
dense inner part of the ONC, also known as the Trapezium Cluster (TC), is characterized by $R_{\rm{TC}}$\,$\lesssim$\,0.3\,pc and
$N_{\rm{TC}}$\,$\approx$\,500, or $\rho_{\rm{TC}}$\,$\approx$\,$4\cdot10^3$\,pc$^{-3}$.

In the most detailed study on circumstellar discs in the Trapezium Cluster, \citet{2000AJ....120.3162L} found a fraction of 80-85\,\% discs among the
stellar population from the $L$-band excess. This is in agreement with an earlier investigation of the complete ONC in which
\citet{1998AJ....116.1816H} report a disc fraction of 50-90\,\% (though relying only on $I_{\mathrm{C}}-K$ colors) and justifies the assumption of a
100\,\% primordial disc fraction in the here presented simulations.

In the following we describe the construction of the numerical cluster models that have been used in our simulations.

%

\section{Computational method}

\label{sec:numerical_method}

The basic dynamical model of the ONC used here is described in \citet{2006ApJ...642.1140O}, with several extensions discussed in
\citet{2007A&A...475..875P}. We summarize the main aspects of our model: The initial stellar population consists of 4000 members with masses between
$0.08\,\Msun$ and $50\,\Msun$ sampled from the standard Kroupa IMF (see Eq.~(\ref{eq:kroupa_imf})). The system is initially in virial equilibrium,
\begin{equation}
  \label{eqn:vir}
  Q_{\rm{vir}}=\frac{R_{\rm{hm}}\sigma^2}{2GM} = 0.5 \,,
\end{equation}
where $R_{\rm{hm}}$ is the half-mass radius of the cluster, $\sigma$ the velocity dispersion, $M$ the total mass, and $G$ the gravitational
constant. It is characterized by a radial density profile, $\rho \propto r^{-2}$, with a central density $\sim$4$\cdot10^4$\,pc$^{-3}$. We adopt a
Maxwellian velocity distribution as would be expected from theory of star cluster formation \citep[e.g.][]{2000prpl.conf..151C} and roughly in
agreement with observations of star-forming regions \cite[e.g.][]{1990ApJ...359..344F,1995ApJ...450L..27M}. We use here a single star model only and
do not include the effect of gas expulsion or stellar evolution. All simulations have been performed with the direct $N$-body code $\nbodypp$
\citep{1999JCoAM.109..407S,2003grav.book.....A}.

For the generation of star cluster models in the present investigation the initial radial density profile has been modified slightly. To first order
the isothermal sphere, $\rho(r) \propto r^{-2}$, represents the projected density distribution of the ONC, yet a flattening in the core, $\Sigma(r)
\propto r^{-0.5}$, $0 < r \le R_\text{core}$, $R_\text{core} \approx 0.2$\,pc, is observed \citep{2005MNRAS.358..742S}. Validating the initial setup
by means of the best reproduction of the \emph{current projected} density distribution of the ONC after a simulation time of 1\,Myr, the evaluation of
numerous initial configurations led to the following best estimate of the \emph{initial three-dimensional} density distribution:
\begin{align}
  \label{eq:density_distribution_initial}
  \rho_0(r) & =
  \begin{cases}
    \rho_0\,(r/R_{\mathrm{core}})^{-2.3} & , \quad r \in (0,R_{\mathrm{core}}] \\
    \rho_0\,(r/R_{\mathrm{core}})^{-2.0} & , \quad r \in (R_{\mathrm{core}},R] \\
    \qquad 0 & , \quad r \in (R,\infty] \\
  \end{cases} \quad ,
\end{align}
where $\rho_0 = 3.1 \cdot 10^3\,\text{stars}\,\text{pc}^{-3}$, $R_{\mathrm{core}} = 0.2\,\text{pc}$, and $R = 2.5\,\text{pc}$.

Moreover, the generation of the high-mass end of the mass function has been modified. In the case of the ONC the upper mass was chosen to be
50\,\Msun\ because this value corresponds to the mass of the most massive stellar system in the ONC. However, stars with larger masses are expected to
form in more massive clusters \citep{2005ApJ...620L..43O,2006MNRAS.365.1333W}. Thus in the framework of this numerical investigation the upper mass
limit has been set to the current accepted fundamental upper mass limit, $m_{\mathrm{max}} = 150\,\Msun$
\citep{2005Natur.434..192F,2005ApJ...620L..43O,2006MNRAS.365..590K,2006MNRAS.365.1333W,2007ApJ...660.1480M,2007ARA&A..45..481Z}.

The choice of a fixed upper mass limit, though in disagreement with the well-established \emph{non-trivial} correlation of the mass of a star cluster
and its most massive member \citep[e.g.][]{2003ASPC..287...65L,2006MNRAS.365.1333W,2008MNRAS.391..711M}, was motivated by the fact that the exact
relation is not known. However, a comparison with the ``sorted sampling algorithm'' of \citet{2006MNRAS.365.1333W} in
Table~\ref{tab:cluster_parameters} shows that -- at least in a statistical sense -- the exact prescription for the generation of the maximum stellar
mass in a cluster is not as important as it might seem. The values obtained by random sampling are only slightly larger than those from sorted
sampling.

\begin{table*}
  \centering
  \begin{tabular}{||c|*2{*{4}{c}|}{|*{3}{c}||}}
    \hline
    \hline
    family of models  &  \multicolumn{4}{c|}{density-scaled}  &  \multicolumn{4}{c||}{size-scaled} &  \multicolumn{3}{c||}{} \\
    \hline
    \hline
    $N$
    & model & $R$  & $\rho_{\mathrm{TC}}$ & $\sigma_{\mathrm{3D}}$
    & model & $R$  & $\rho_{\mathrm{TC}}$ & $\sigma_{\mathrm{3D}}$ 
    & $m_{\mathrm{max}}^{\mathrm{med}}$ & $m_{\mathrm{max}}^{\mathrm{sort}}$ & $m_{\mathrm{max}}^{\mathrm{obs}}$  \\

    &       & [pc] & [$10^3$\,pc$^{-3}$]  & [\kms]
    &       & [pc] & [$10^3$\,pc$^{-3}$]  & [\kms]             
    & [\Msun]                       & [\Msun]                        & [\Msun]  \\
    \hline
    ~~1000  &  D0  &  ~~2.50  &  ~~1.3  &  1.15  &  S0  &  ~~0.63  &  ~~4.8  &  2.42  &  ~~36  &  ~~32  &  $25 \pm 15$  \\
    ~~2000  &  D1  &  ~~2.50  &  ~~2.7  &  1.64  &  S1  &  ~~1.25  &  ~~5.1  &  2.37  &  ~~52  &  ~~45  &  $25 \pm 15$  \\
    ~~4000  &  D2  &  ~~2.50  &  ~~5.3  &  2.26  &  S2  &  ~~2.50  &  ~~5.3  &  2.26  &  ~~79  &  ~~63  &  $55 \pm 25$  \\
    ~~8000  &  D3  &  ~~2.50  &   10.5  &  3.11  &  S3  &  ~~5.00  &  ~~5.3  &  2.13  &  ~~94  &  ~~80  &  $75 \pm 25$  \\
    16000  &  D4  &  ~~2.50  &   21.1  &  4.34  &  S4  &   10.00  &  ~~5.3  &  2.11  &   125  &   112  &  $95 \pm 35$  \\
    32000  &  D5  &  ~~2.50  &   42.1  &  6.04  &  S5  &   20.00  &  ~~5.2  &  2.03  &   137  &   126  &  $100\pm 35$  \\
    \hline
    \hline
  \end{tabular}
  \caption{Averaged initial parameters of the cluster models, divided among the families of density-scaled and size-scaled models. The first column
    contains the number of stars, $N$, the next eight columns contain the designation, the cluster size, $R$, the number density in a sphere of radius 0.3\,pc,
    $\rho_{\mathrm{TC}}$ (equivalent to the Trapezium Cluster in the ONC, see Section~\ref{sec:observations_onc}), and the
    three-dimensional velocity dispersion, $\sigma_{\mathrm{3D}}$, for the density-scaled and size-scaled models, respectively. The last three columns
    denote the median maximum stellar mass in each simulation, $m_{\mathrm{max}}^{\mathrm{med}}$, the median maximum stellar mass for sorted
    sampling, $m_{\mathrm{max}}^{\mathrm{sort}}$, and estimates from observational data, $m_{\mathrm{max}}^{\mathrm{obs}}$, both taken from Fig.~7 of
    \citet{2006MNRAS.365.1333W}.}
  \label{tab:cluster_parameters}
\end{table*}

Stellar encounters in dense clusters can lead to significant transport of mass and angular momentum in protoplanetary discs
\citep{2006ApJ...642.1140O,2006A&A...454..811P,2007A&A...462..193P}. In the present investigation we assume that all discs are of low mass, i.e. a low
mass ratio of disc and central star, $m_{\mathrm{disc}}/m_{\star} \ll 0.1$, in agreement with observations of the ONC
\citep{1998AJ....116..854B,2005ApJ...634..495W,2009ApJ...699L..55M}. We use Eq.~(1) from \citet{2006A&A...454..811P} to keep track of the
\emph{relative} disc-mass loss of each star-disc system due to encounters. Approaches of stars are only considered to be encounters if the calculated
relative disk-mass loss is higher than 3\,\%, corresponding to the 1$\sigma$ error in our simulations of star-disc encounters. In this case the
relative disc-mass loss is independent of the disc mass and depends only on the mass ratio of the interacting stars and the orbital parameters
\citep[see][]{2006A&A...454..811P}. Our estimate of the accumulated disc-mass loss is an upper limit because the underlying formula is only valid for
co-planar, prograde encounters, which are the most perturbing. A simplified prescription assigns stars into one of two distinct groups: if the
relative disc-mass loss exceeds 90\,\% of the initial disc mass, stars are marked as ``discless''; otherwise they are termed ``star-disc systems''. As
in our previous investigations, the determination of the disc-mass loss involves two different models of the initial distribution of disc sizes:
i)~scaled disc sizes $r_{\mathrm{d}}$ with $r_{\mathrm{d}} = 150\,\mathrm{AU} \sqrt{M_1} [\Msun]$, which is equivalent to the assumption of a fixed
force at the disc boundary, and ii)~equal disc sizes with $r_{\mathrm{d}} = 150\,\mathrm{AU}$. Whenever results are presented, we will specify which
of these two distributions has been used.

%
\section{Numerical models}

\label{sec:numerical_models}

For the present study we have set up a variety of scaled versions of the standard ONC model cluster with varying sizes and stellar numbers. In total,
eleven cluster models have been set up (see Table~\ref{tab:cluster_parameters}). They form two parametric groups, the ``density-scaled'' (D0-D5) and
the ``size-scaled'' (S0-S5) group, both containing six clusters with stellar numbers of 1000, 2000, 4000, 8000, 16000, and 32000, respectively. Models
D2 and S2 are identical and correspond to the standard ONC model with the adopted higher stellar upper mass limit, $m_{\mathrm{max}} =
150\,\Msun$. The other ten cluster models have been set up as scaled representations of ONC-like clusters. As in the case of the numerical model of
the ONC, for each cluster model a set of simulations has been performed with varying random configurations of positions, velocities, and masses,
according to the given distributions, to lower the effect of statistical uncertainties. For the clusters with 1000, 2000, 4000, 8000, 16000, and 32000
particles, a number of 200, 100, 100, 50, 20, and 20 simulations seemed appropriate to provide sufficiently robust results.

It has to be noted that these ``artificial'' stellar systems are not just theoretical models but have as well counterparts in the observational
catalogues of star clusters: the young star cluster NGC~2024 \citep[e.g.][]{2000AJ....120.1396H,2003A&A...404..249B,2003AJ....126.1665L} is well
represented by the 1000 particle model, whereas the 16000 particle model has its counterpart in the massive cluster NGC~3603
\citep[e.g.][]{1999AJ....117.2902D,2004AJ....128..765S,2006AJ....132..253S}.

\subsubsection*{Density-scaled cluster models}
The six density-scaled cluster models (D0-D5) have been simulated with the same initial size as the ONC ($R=2.5$\,pc). Due to the adopted number
density distribution, roughly represented by $\rho(r) = \rho_0 \, r^{-2}$, the density of the models scales as the stellar number,
\begin{equation}
  \label{eq:relation_number_density_size}
  N = \int_0^R \rho(r) \, r^2 \, dr \, d\Omega \propto \rho_0 \, R \,,
\end{equation}
though for an exact treatment one would have to consider the steeper density profile of the core, $\rho_{\mathrm{core}}(r) = \rho_{\mathrm{core},0} \,
r^{-2.3}$,
\begin{equation}
  N_{\mathrm{core}} = \int_0^{R_{\mathrm{core}}} \rho_{\mathrm{core}}(r) \, r^2 \, dr \, d\Omega \propto \rho_{\mathrm{core},0} \; R_{\mathrm{core}}^{0.7} \,.
  \label{eq:relation_number_density_size_core}
\end{equation}
However, since the core population is not dominant in terms of number, the clusters are characterised in good approximation by densities that are 1/4,
1/2, 1, 2, 4, and 8 times the density of the ONC (at any radius), respectively. These models are used to study the importance of the density for the
effect of star-disc encounters in a cluster environment.

\subsubsection*{Size-scaled cluster models}
Five more cluster models have been simulated with the same initial density as the ONC but with varying extension. The set of size-scaled cluster
models (S0-S5) is used to study the pure effect of the size of the stellar population. Due to the relation~(\ref{eq:relation_number_density_size}),
the initial size of these clusters scales as the stellar number and was set up with 1/4, 1/2, 1, 2, 4 and 8 times the initial size of the ONC,
respectively.

\bigskip

The initial parameters of the cluster models, for each model averaged over all configurations, are presented in
Table~\ref{tab:cluster_parameters}. Here the number density in the Trapezium Cluster, $\rho_{\mathrm{TC}}$, is taken as a reference value for all
simulations. As expected, the density scales with the number of stars for the density-scaled models, while it is rather constant for the size-scaled
models. The velocity dispersion, that satisfies the relation
\begin{equation}
  \label{eq:relation_velocity_dispersion}
  \sigma = \sqrt{ \frac{2GM}{R} } \propto \sqrt{ \frac{N}{R} } \,,
\end{equation}
shows the expected scaling of $\sqrt{N}$ for the density-scaled models, and is again roughly constant for the size-scaled models, as expected from $N
\propto R$ (Eq.~(\ref{eq:relation_number_density_size})) and the above relation. The reason for the slight increase of the velocity dispersion with
decreasing stellar number for the size-scaled models is the steeper density profile in the cluster core, which becomes more dominant in terms of
stellar number with decreasing cluster size. Combining $N_{\mathrm{core}} \propto R^{0.7}$ from Eq.~(\ref{eq:relation_number_density_size_core}) and
the above relation gives roughly $\sigma_{\mathrm{core}} \propto N^{-0.3}$ and thus explains the correlation.

%

\section{Results of numerical simulations}

\label{sec:numerical_results}

In this section the results of the numerical simulations of the two families of cluster models will be presented. A short discussion of the
characteristic scaling relations and differences of the cluster dynamics will be followed by a more detailed investigation of the encounter-induced
disc-mass loss among the two families of density-scaled and size-scaled cluster models.

\subsection{Cluster dynamics}

\label{sec:numerical_results:cluster_dynamics}

\begin{figure}
  \centering
  \includegraphics[width=1.0\linewidth]{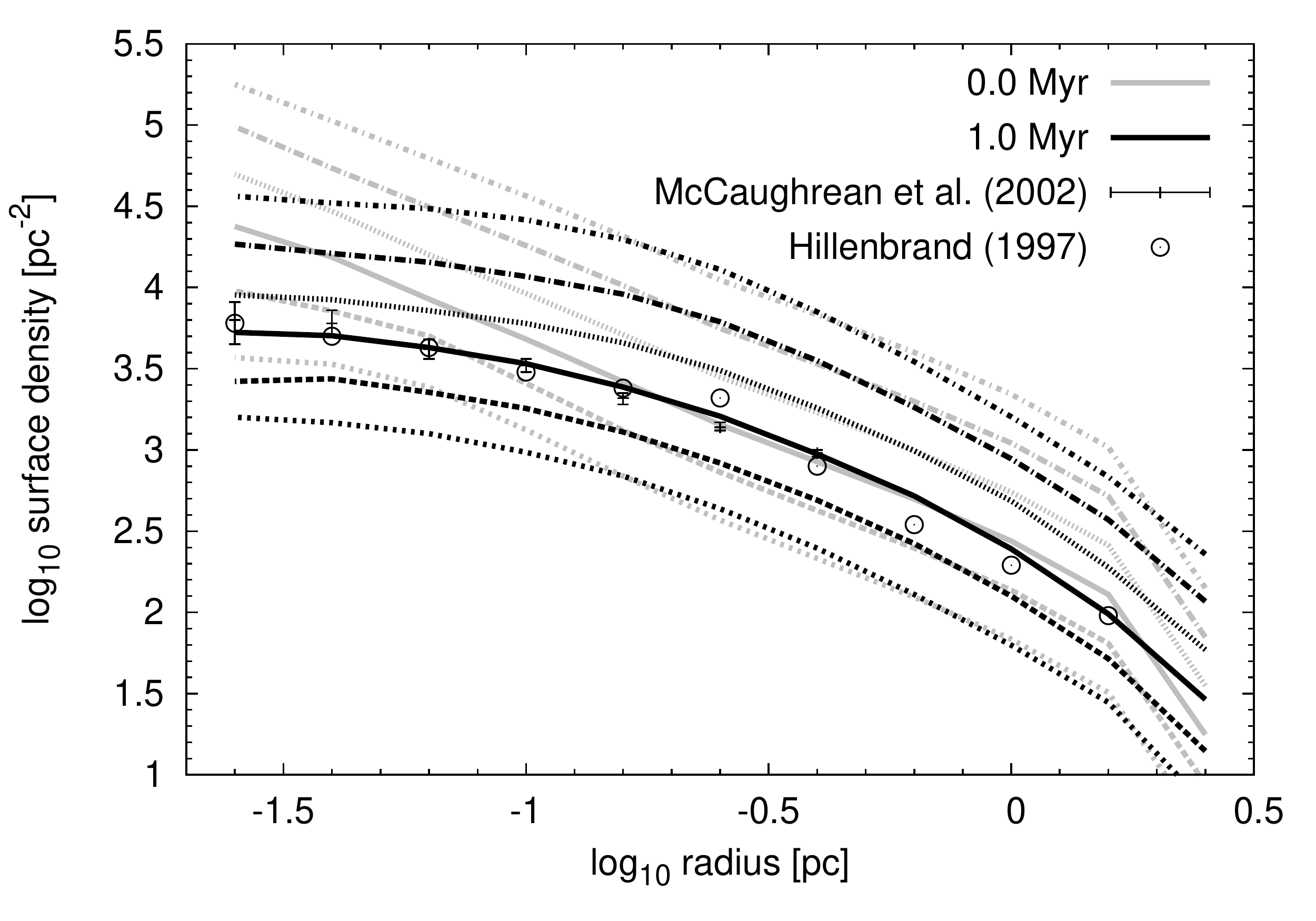}
  \caption{Projected density profiles of the density-scaled models compared to observational data. The initial profile (grey lines) and the profile at
    a simulation time of 1\,Myr (black lines) are shown. From bottom to top in each colour regime the cluster models D0-D5 are marked by a
    short-dashed, long-dashed, solid, dotted, dot-long-dashed, and dot-short-dashed line, respectively. The observational data are from a compilation
    of \citet{2002Msngr.109...28M} and \citet{1997AJ....113.1733H}.}
  \label{fig:projected_density_distribution_sample_density_scaled}
\end{figure}

In Fig. \ref{fig:projected_density_distribution_sample_density_scaled} the evolution of the projected density distribution of the density-scaled
models is shown. The shape of the distributions is in all cases very similar. The evolved distributions have nearly identical shapes and are
separated by vertical intervals of 0.3 in log-space, which corresponds to the difference of the initial densities by a factor~2. Only in the innermost
cluster regions slight deviations between the evolved distributions are apparent. These are attributed to the poorer random sampling of the initial
particle distribution due to the very steep density profile, $\rho(r) \propto r^{-2.3}$, as is evident from the larger scatter among the initial
profiles. However, after 1\,Myr these deviations are smoothed out to a large degree.

Due to the nearly exact qualitative \emph{and} scaled quantitative evolution of the density-scaled cluster models, it is justified to ascribe
differences of the effects of star-disc encounters on the stellar population mainly to one parameter, namely the initial density of the cluster
models. Nevertheless, the influence of the particle number has to be considered, too.

\begin{figure}
  \centering
  \includegraphics[width=1.0\linewidth]{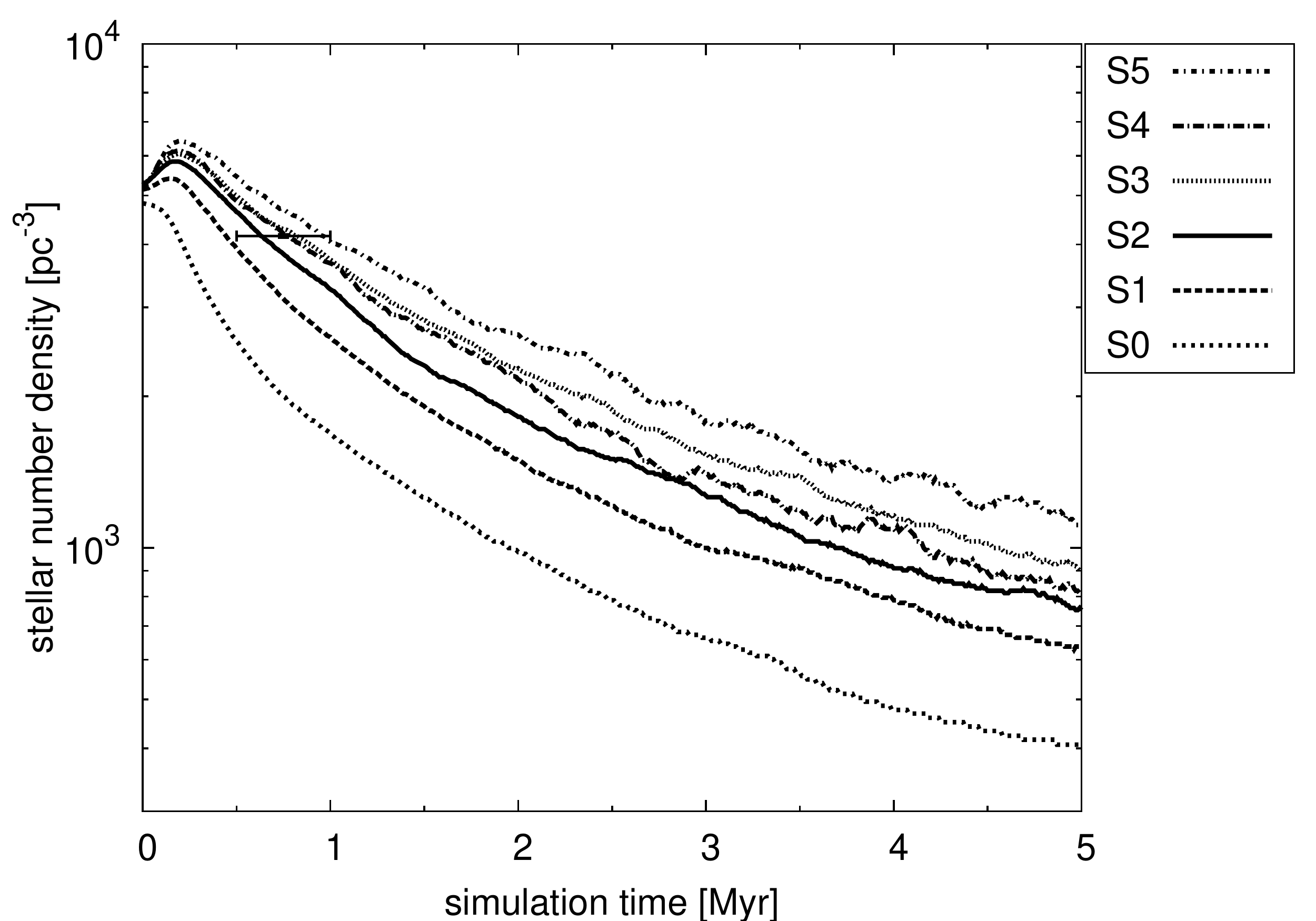}
  \caption{Time evolution of stellar densities $\rho_{\mathrm{TC}}$ (in a volume of radius $R_{\mathrm{TC}}=0.3\,$pc, see
    Section~\ref{sec:observations_onc}) of the size-scaled cluster models. The cluster models S0-S5 are marked by a short-dashed, long-dashed, solid,
    dotted, dot-long-dashed, and dot-short-dashed line, respectively. The horizontal error bar marks the corresponding observational estimates for
    comparison.}
  \label{fig:density_vs_time_sample_size_scaled}
\end{figure}

The size-scaled cluster models show a different dynamical evolution compared to the density-scaled models. The temporal evolution of the densities
$\rho_{\mathrm{TC}}$ in Fig.~\ref{fig:density_vs_time_sample_size_scaled} demonstrates that the clusters evolve on slightly different time scales,
where the density declines faster for the less populated clusters S0 and S1. However, the densities of the models S1-S5 differ not much, and are
consistent with a coeval decline. The evolution of the cluster densities does not -- to first order -- depend on the number of particles, probably
with the exception of the model S0. The size-scaled models are thus well suited to investigate the effect of the number of cluster stars on star-disc
encounters and the corresponding induced disc-mass loss.

\subsection{Encounter dynamics}

\label{sec:numerical_results:encounter_dynamics}

\begin{figure}
  \centering
  \includegraphics[width=1.0\linewidth]{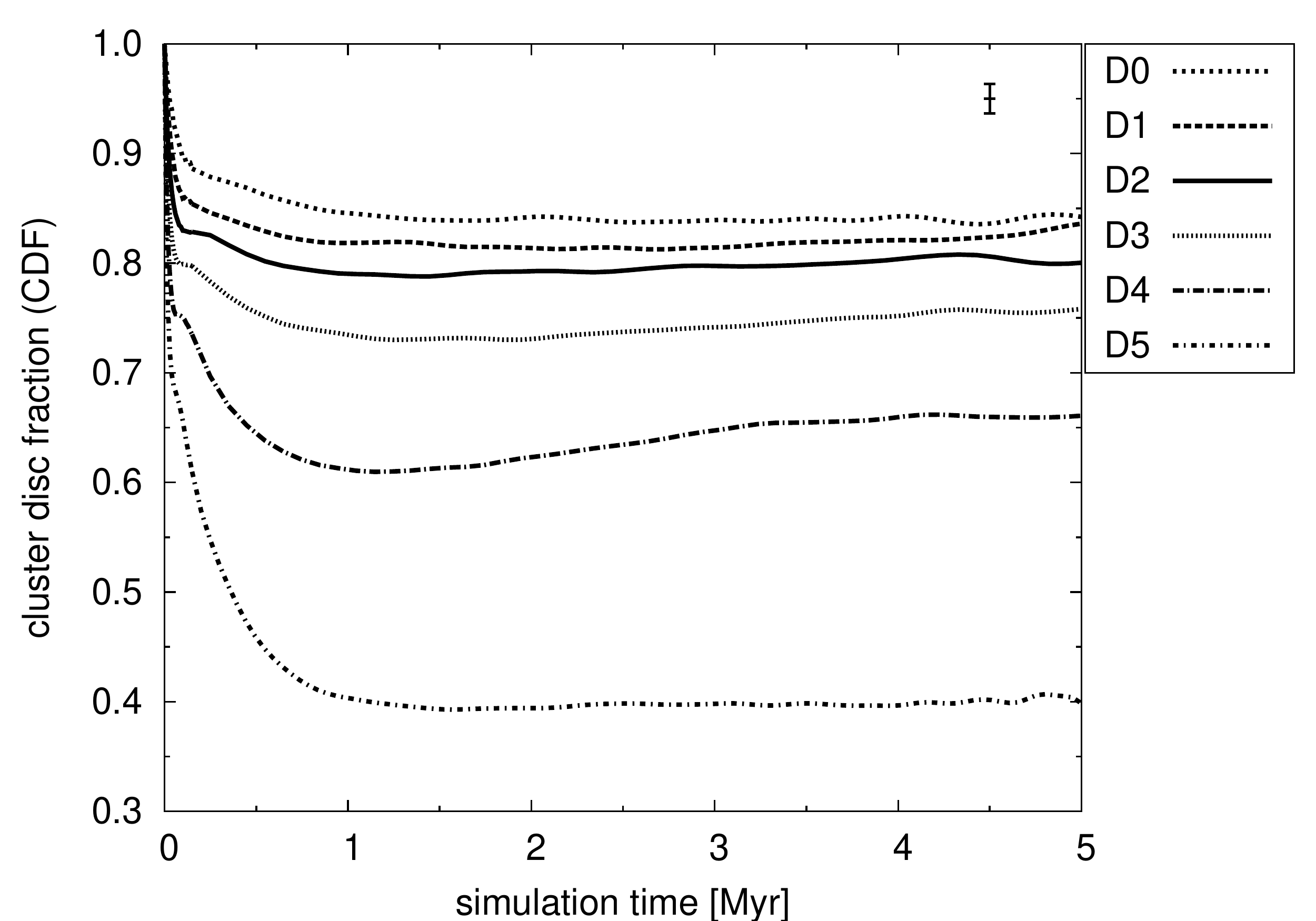}
  \caption{Time evolution of the cluster disc fraction of the density-scaled models for a region of the size of the Trapezium Cluster
    ($R_{\mathrm{TC}}=0.3\,\mathrm{pc}$). The curves have been smoothed by Bezier curves to avoid intersecting lines. Here initially equal disc sizes
    have been assumed. From top to bottom the cluster models D0-D5 are marked by a short-dashed, long-dashed, solid, dotted, dot-long-dashed, and
    dot-short-dashed line, respectively. The typical error bar is indicated in the upper right.}
  \label{fig:disc_destruction_vs_time__sample_density_scaled}
\end{figure}

We investigate the effect of the cluster density on the encounter-induced disc-mass loss via the evolution of the cluster disc fraction (CDF). The
distributions in Fig.~\ref{fig:disc_destruction_vs_time__sample_density_scaled} show the average fraction of stars that are surrounded by disc
material and correspond from top to bottom to clusters with increasing density. The curves have been smoothed with Bezier curves to provide a clearer
view. It is evident that the fraction of destroyed discs in the Trapezium Cluster increases significantly with increasing cluster density. In
particular, the effect becomes much stronger for the models D3-D5 with 2-8 times the density of the ONC. In the case of the densest model D5, even up
to 60\,\% of the stars in the Trapezium Cluster could have lost their discs after 1\,Myr of dynamical evolution. But it is also interesting to note
that even in a cluster 4 times less dense than the ONC (model D0), still 10-15\,\% of the stars could lose their surrounding discs due to
gravitational interactions with cluster members. However, one has to treat these numbers with care due to the -- partially -- significant
uncertainties that go into the calculations \citep[see][]{2006ApJ...642.1140O}. Nonetheless, what is more important here -- and relies only on the
\emph{relative} quantities -- is the fact that the distributions in Fig.~\ref{fig:disc_destruction_vs_time__sample_density_scaled} are not equidistant
but do show larger differences with increasing density.

The impression that the disc fraction increases at later stages of the cluster evolution is solely due to the acceleration of perturbed systems
leaving the sampling volume. The escape rate follows the trend of the disc destruction rate with a time delay that is determined by the difference of
the crossing time of disc-less stars and star-disc systems in the sampling volume. Hence, the decrease of the disc fraction is followed by an increase
and eventually remains constant some time after disc destruction has stopped.

The preferred mass range of stars that become disproportionately high involved in perturbing encounters changes significantly between the models D2
and D4. From previous investigations \citep{2006ApJ...642.1140O,2006A&A...454..811P} we already know that in the model D2, representing an ONC-like
cluster, the high-mass stars' discs are mostly affected by encounters. These encounters occur preferentially with low-mass stars due to gravitational
focusing. However, in the model cluster D4, a four times denser system, it is the low-mass stars that dominate the disc-mass loss of the cluster
population.

\begin{figure}
  \centering
  \includegraphics[width=1.0\linewidth]{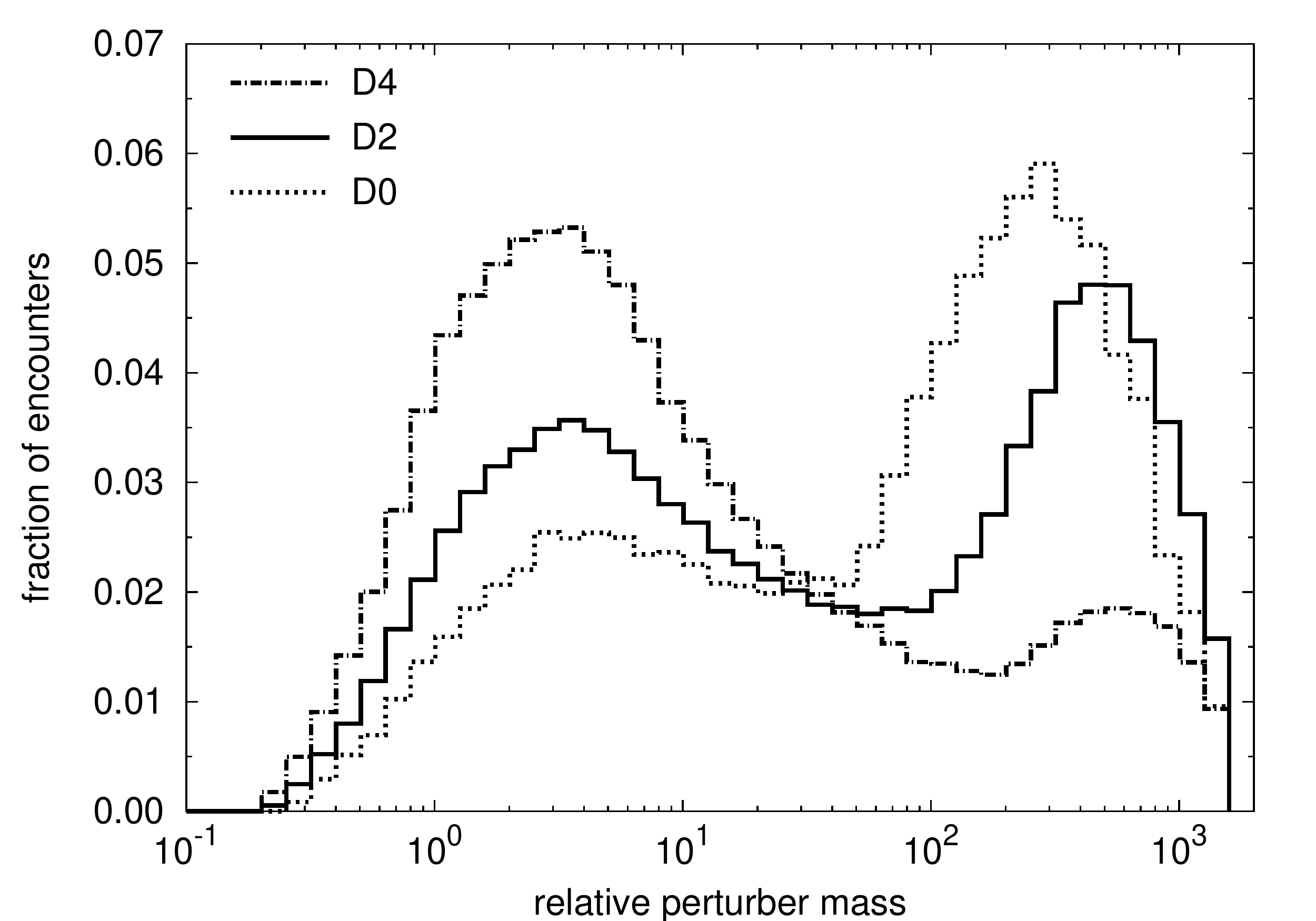}
  \caption{Comparison of the fraction of encounters as a function of the relative perturber mass, i.e. the mass ratio of perturber and perturbed star,
    of the group of low-mass stars (see Appendix~\ref{app:star_cluster_models:mass_groups} for the width of the mass intervals) for the models D0
    (dotted), D2 (solid), and D4 (dot-long-dashed), respectively. Here initially equal disc sizes have been assumed.}
  \label{fig:encounters_vs_relative_perturber_mass__disc_mass_loss__lowmass__4000_1000_16000__sample_density_scaled}
\end{figure}

\begin{figure}
  \centering
  \includegraphics[width=1.0\linewidth]{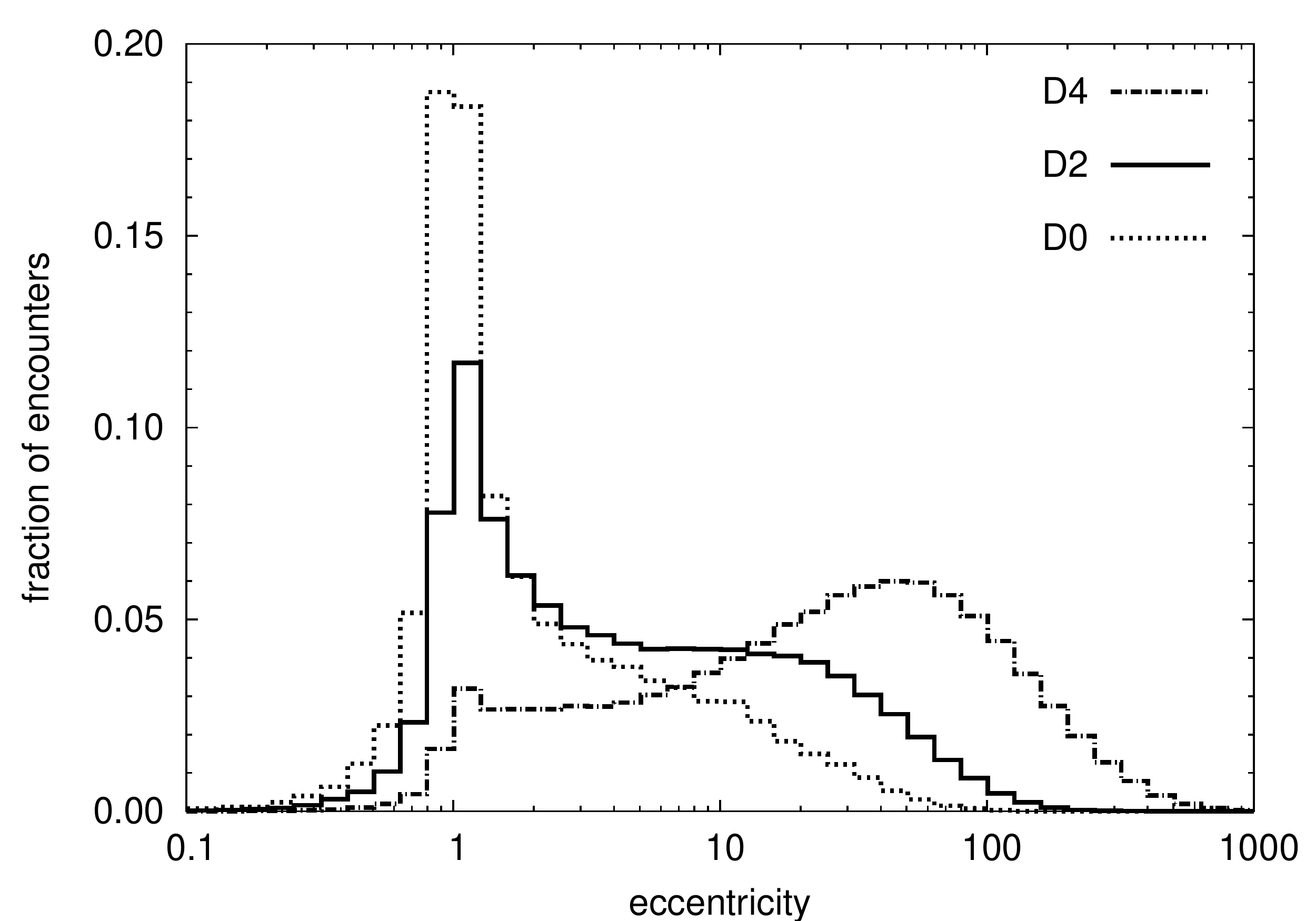}
  \caption{Comparison of the fraction of encounters as a function of eccentricity of the group of low-mass stars (see
    Appendix~\ref{app:star_cluster_models:mass_groups} for the width of the mass intervals) for the models D0 (dotted), D2 (solid), and D4
    (dot-long-dashed), respectively. Here initially equal disc sizes have been assumed.}
  \label{fig:encounters_vs_eccentricity__disc_mass_loss__lowmass__4000_1000_16000__sample_density_scaled}
\end{figure}

We demonstrate this transition by binning the fraction of encounters, normalised to the total number of encounters of each model (so that the
integrated surface is unity in each case). Recall that encounters have been defined as flybys causing more than 3\,\% disk-mass loss (see
Section~\ref{sec:numerical_method}). Fig.~\ref{fig:encounters_vs_relative_perturber_mass__disc_mass_loss__lowmass__4000_1000_16000__sample_density_scaled}
shows for the group of low-mass stars that the dominance of perturbations from much more massive stars in the models D0-D2, represented by the peak at
relative perturber masses $\gtrsim 100$, changes dramatically towards encounters of nearly equal-mass stars in the model D4 (with a peak at relative
perturber masses $\approx 3$). Similarly, the typical geometry of the encounter orbits underlies a strong transition. As is evident from
Fig.~\ref{fig:encounters_vs_eccentricity__disc_mass_loss__lowmass__4000_1000_16000__sample_density_scaled}, most encounters of low-mass stars in the
models D0-D2 are parabolic (with an eccentricity $e \approx 1$). In contrast, in the model D4 these stars usually experience a strongly hyperbolic
flyby of a perturber, mostly with eccentricity $e > 10$. Hence, with increasing cluster density, the dominant mode of star-disc interactions changes
towards hyperbolic encounters of low-mass stars between each other.

This transition is a consequence of gravitational focusing by high-mass stars becoming significantly less important among the models
D0-D5. Gravitational focusing forces a deflection of the interacting stars onto a less eccentric orbit, effectively increasing the cross-section above
geometrical:
\begin{equation}
  \label{eq:cross_section}
  \pi{b}^2 = \pi{r_{\mathrm{min}}^2} (1 + \Theta) \,,
\end{equation}
where
\begin{equation}
  \Theta = \frac{v_{\mathrm{min}}^2}{\langle v_{\mathrm{rel}} \rangle^2} = \frac{\pi}{16}\frac{v_{\mathrm{min}}^2}{\sigma^2}
\end{equation}
is the gravitational focusing term or Safronov number, $b$ the impact parameter, $r_{\mathrm{min}}$ the minimum distance, $v_{\min}$ the escape speed
at the minimum distance, $\langle v_{\mathrm{rel}} \rangle$ the mean relative speed, and $\sigma$ the cluster velocity dispersion (one finds $\langle
v_{\mathrm{rel}} \rangle = 4\sigma / \sqrt{\pi}$ for a Maxwellian distribution with dispersion $\sigma$). Adopting the typical cluster and encounter
parameters from Table~\ref{tab:cluster_parameters} and \ref{tab:encounter_rate_parameter_space}, respectively, and substituting $v_{\mathrm{min}} =
v_{\mathrm{enc}}$ and $r_{\mathrm{min}} = r_{\mathrm{enc}}$, we find that gravitational focusing by low-mass stars is negligible ($\Theta \ll 1$) in
all our cluster models. In contrast, high-mass stars in the model D0 play a substantial role as gravitational foci ($\Theta \approx 120$), whereas
their cross section is reduced by more than one order of magnitude in the model D5 ($\Theta \approx 4$).

In summary, the encounter-induced disc-mass loss in cluster environments of different densities shows two important features: i) low-mass stars lose a
larger fraction of their disc material with increasing cluster density, and ii) the discs of the most massive stars are (nearly) completely destroyed,
independent of the density of the cluster environment. The important finding for i) is that the correlation is not linear, but shows a much larger
increase for the cluster models with higher densities than model D2, implying that there exists a critical density $\rho_{\mathrm{crit}}$ that marks
the onset of a much more destructive effect of star-disc encounters. This critical density seems to be close to the density of the ONC,
$\rho_{\mathrm{crit}} \approx \rho_{\mathrm{ONC}}$.

We find that the evolution of the CDF in the Trapezium Cluster region is very similar among the size-scaled models and corresponds to the distribution
of the model D2 in Fig.~\ref{fig:disc_destruction_vs_time__sample_density_scaled}. The size-scaled models are obviously equivalent in their
environmental effect on protoplanetary discs despite the slightly different dynamical evolution. The density of the models S0 and S1 decreases faster
than for the more massive clusters, even up to a factor of 2 in case of the model S0 (see Section~\ref{sec:numerical_results:cluster_dynamics}). Thus
one would expect a lower encounter rate in these smaller systems and, consequently, on average a lower disc-mass loss. However, this is obviously not
the case, and is explained by the fact that, similarly to the finding for the density-scaled models, high-mass stars become less important as
gravitational foci for the low-mass stars in clusters with larger stellar populations. Hence in terms of encounter statistics, the lower density of
the models S0 and S1 is compensated by the more frequent interactions of the high-mass stars.

\subsection{Validation of the numerical method}

\label{sec:numerical_results:validation}

So far the disc-mass loss has been calculated from Eq.~(1) of \citet{2006A&A...454..811P} by treating all encounters as parabolic. To account for the
lower disc-mass loss in hyperbolic encounters, we have determined a function that quantifies the reduction of the disc-mass loss in dependence of the
orbital eccentricity:
\begin{equation}
  \label{eq:fit_function_mass_loss_vs_eccentricity}
  \begin{split}
    \widehat{\Delta{m}}(e) =& \exp[0.12(e - 1)] \, \times \\
    & \{0.83 - 0.015(e - 1) + 0.17 \exp[0.1(e - 1)]\} \,.
  \end{split}
\end{equation}
Eq.~(\ref{eq:fit_function_mass_loss_vs_eccentricity}) is a fit function to the median distribution of the relative disc-mass loss as a function of
eccentricity, normalised to the parabolic case, for all star-disc simulations that have been performed.

\begin{figure}
  \centering
  \includegraphics[width=1.0\linewidth]{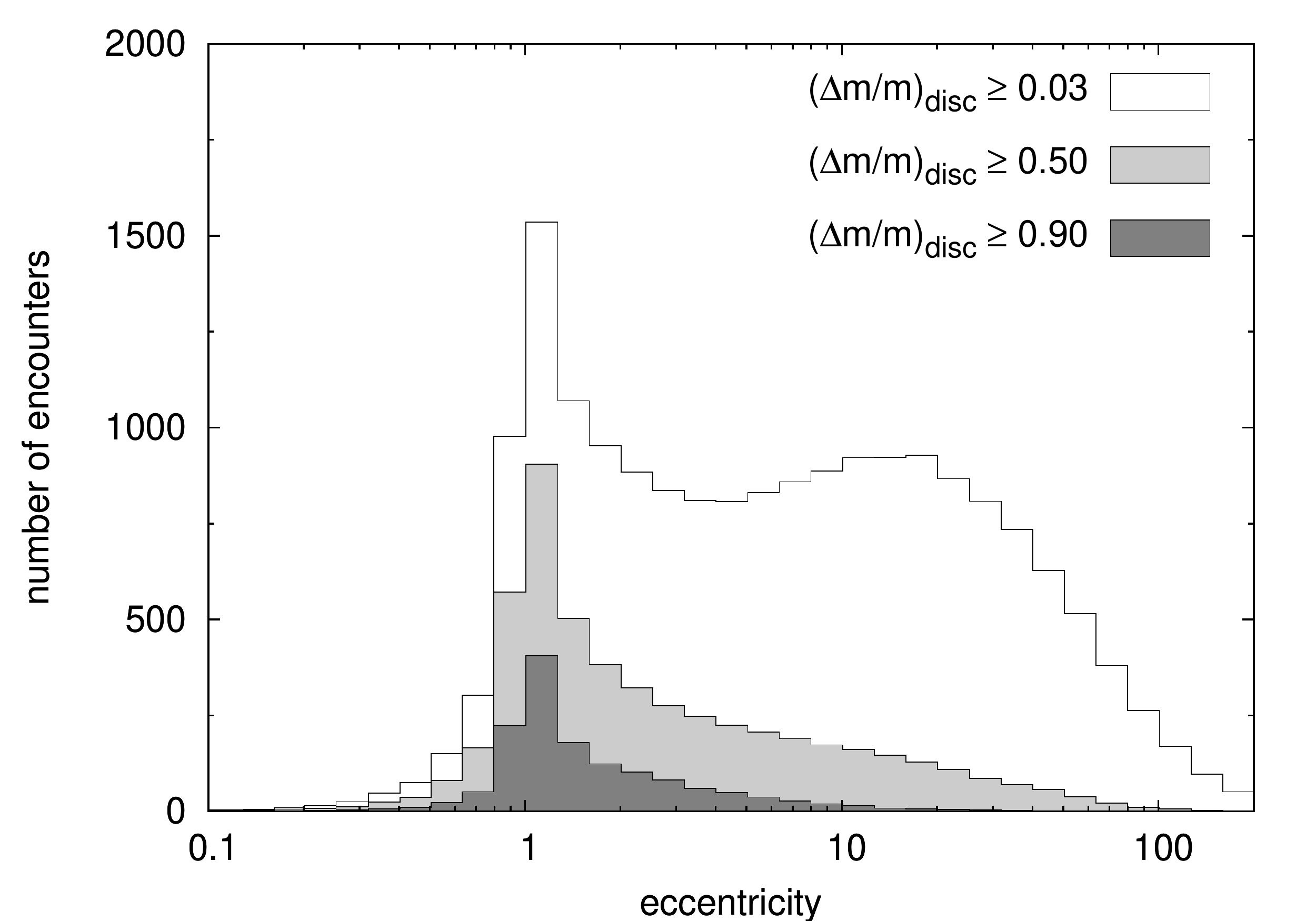}
  \includegraphics[width=1.0\linewidth]{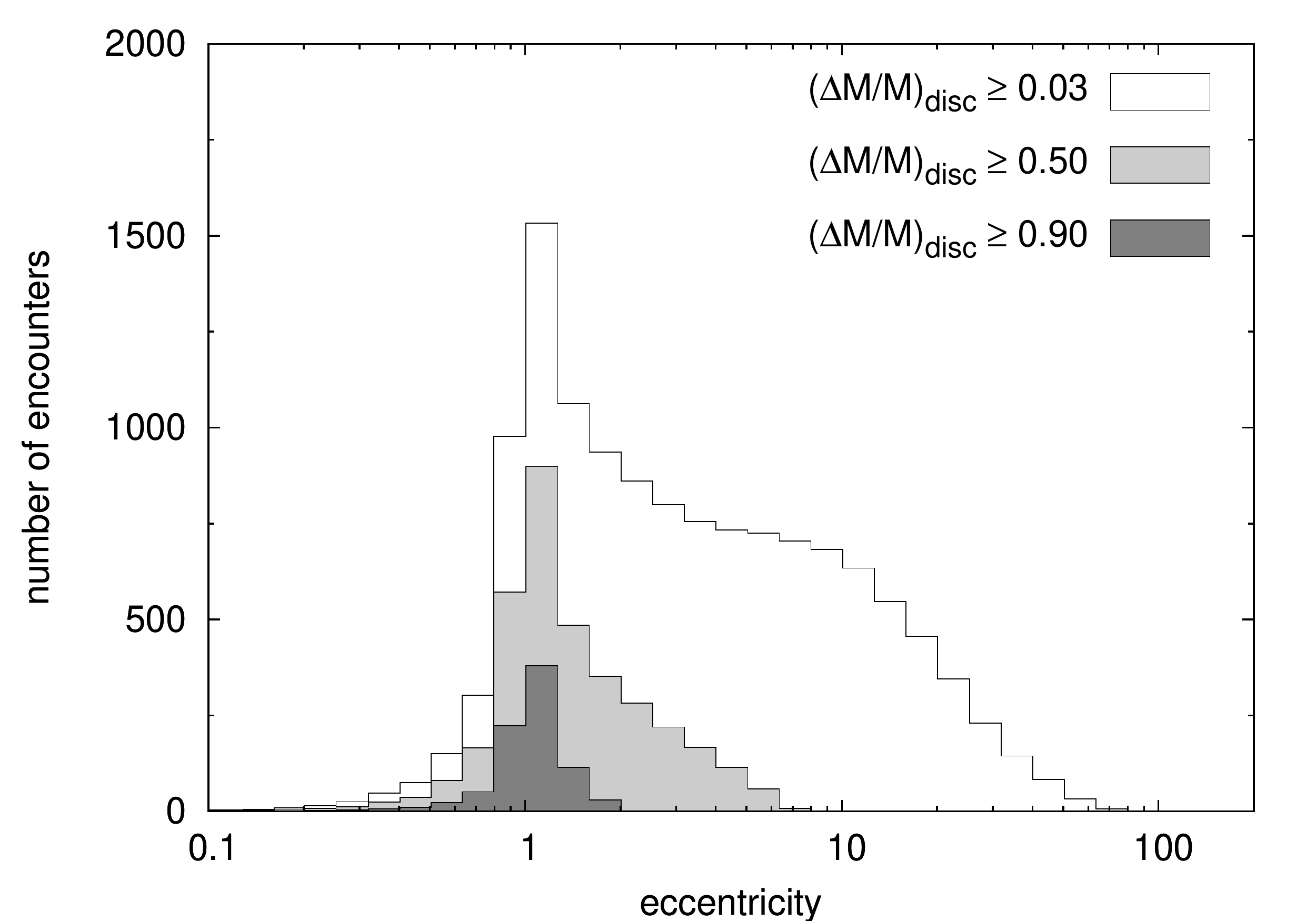}
  \caption{Number of encounters as a function of eccentricity (logarithmic bins), plotted for three different groups by means of disc-mass loss per
    encounter. The white surface represents all encounters \citep[i.e. minimum 3\,\% disc-mass loss, cf.][]{2006ApJ...642.1140O}, the light grey those
    that removed at least 50\,\% of the disc-mass, while the dark grey stands for the most destructive encounters that caused a disc-mass loss of at
    least 90\,\%. The two plots show the distributions for the model D4. \emph{Top:} Disc-mass loss calculated assuming parabolic
    encounters. \emph{Bottom:} Disc-mass loss calculation corrected for effects of eccentricity by using
    Eq.~(\ref{eq:fit_function_mass_loss_vs_eccentricity}).}.
  \label{fig:encounters_vs_eccentricity__disc_mass_loss__parabolic_and_ecc_model__sample_density_scaled_n16000}
\end{figure}

As shown in Fig.~\ref{fig:encounters_vs_eccentricity__disc_mass_loss__parabolic_and_ecc_model__sample_density_scaled_n16000}, the number of hyperbolic
encounters is significantly reduced if the eccentricity is considered explicitly in the calculation of the disc-mass loss using the fit function
(\ref{eq:fit_function_mass_loss_vs_eccentricity}). This is a consequence of our definition of an encounter (see Section~\ref{sec:numerical_method}):
the fraction of perturbations that cause a relative disc-mass loss above 3\,\% is lower for higher eccentricities, hence the number of eccentric
encounters decreases. However, because the fit function represents the \emph{median} distribution of all simulated star-disc encounters and the effect
of strongly perturbing encounters is only weakly dependent on the eccentricity, the number of strongly perturbing encounters in
Fig.~\ref{fig:encounters_vs_eccentricity__disc_mass_loss__parabolic_and_ecc_model__sample_density_scaled_n16000}b (light and dark grey surfaces) is
underestimated.

\begin{figure}
  \centering
  \includegraphics[width=1.0\linewidth]{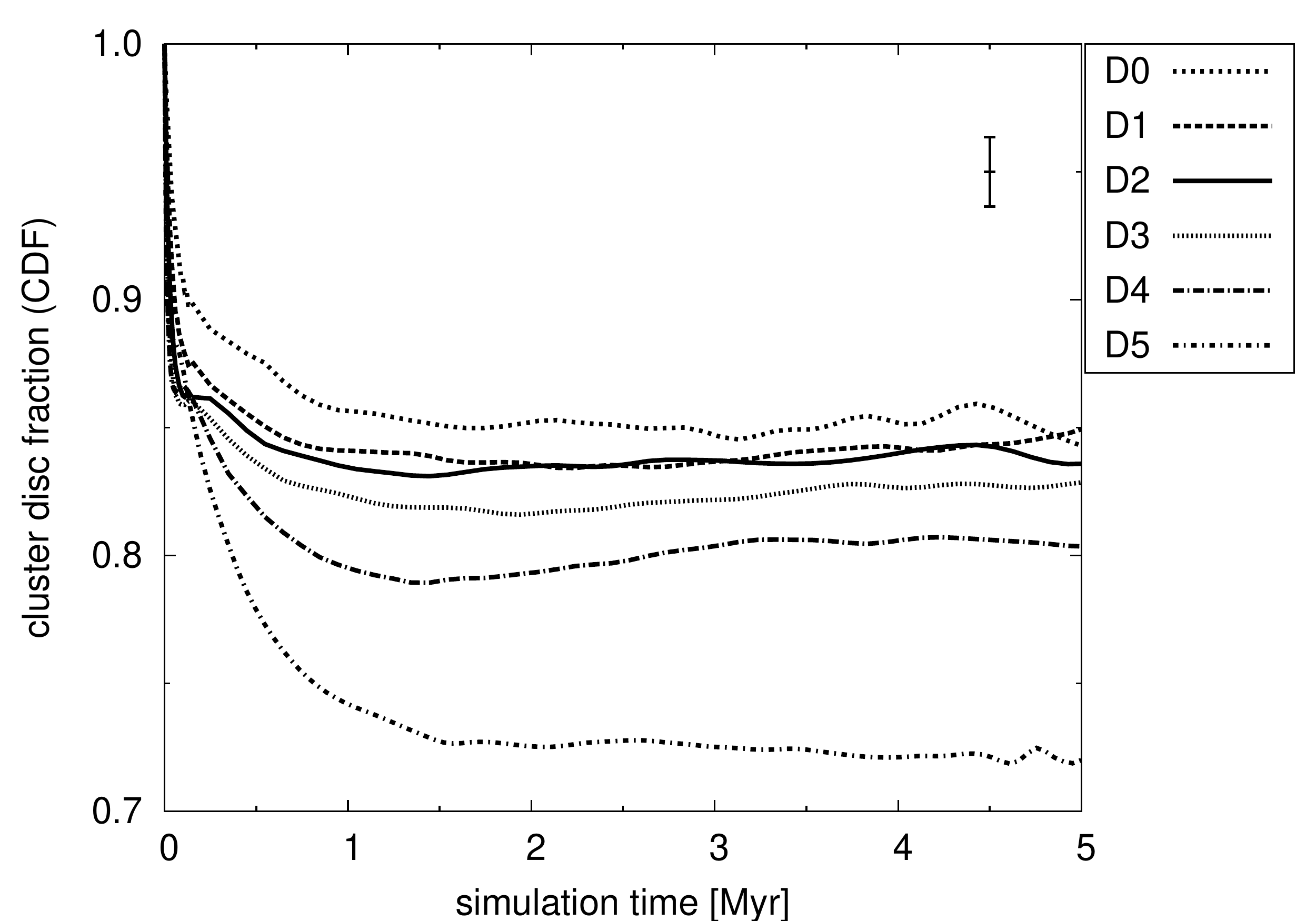}
  \caption{Time evolution of the cluster disc fraction of the density-scaled models, not restricted to parabolic encounters, for a region of the size
    of the Trapezium Cluster ($R_{\mathrm{TC}}=0.3\,\mathrm{pc}$). The curves have been smoothed by Bezier curves to avoid intersecting lines. Here
    initially equal disc sizes have been assumed. From top to bottom the cluster models D0-D5 are marked by a short-dashed, long-dashed, solid,
    dotted, dot-long-dashed, and dot-short-dashed line, respectively. The typical error bar is indicated in the upper right.}
  \label{fig:disc_destruction_vs_time__ecc_model__sample_density_scaled}
\end{figure}

Hence it is the disc-mass loss induced by weak hyperbolic interactions that has been overestimated in the calculations. These events are most numerous
in the models D3-D5 with densities $\rho > \rho_{\mathrm{crit}}$ and result preferentially from interactions of low-mass stars with roughly equal-mass
perturbers. Consequently, when considering explicitly the reduced disc-mass loss in hyperbolic encounters (see
Fig.~\ref{fig:disc_destruction_vs_time__ecc_model__sample_density_scaled}), the outstanding role of these dense clusters as environments of huge disc
destruction becomes less pronounced (cf. Fig.~\ref{fig:disc_destruction_vs_time__sample_density_scaled}), though the encounter-induced disc-mass loss
is still considerably larger compared to the sparser clusters D0-D2.

\bigskip

\begin{figure}
  \centering
  \includegraphics[width=1.0\linewidth]{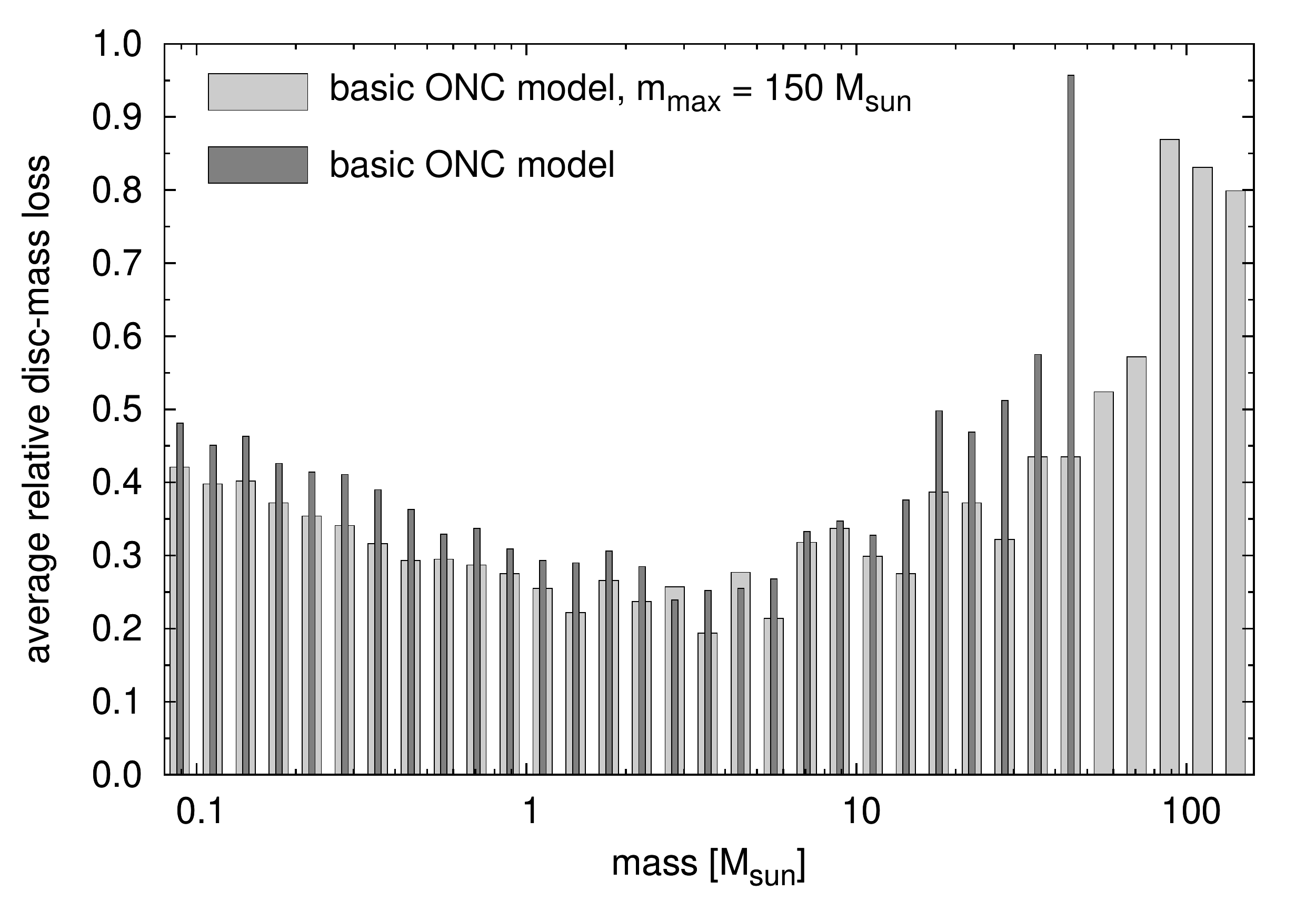}
  \caption{Average relative disc-mass loss at 1\,Myr for the Trapezium Cluster as a function of the stellar mass for initially equal disc sizes. The
    standard ONC model with a stellar upper mass limit of 50\,\Msun\ (dark grey bars) is compared to a simulation of the same model with an upper mass
    limit of 150\,\Msun\ (light grey bars).}
  \label{fig:disc_mass_loss_vs_mass_bins__disc_size_fixed__basic_onc_model_50_msun_and_150_msun}
\end{figure}

The effect of the higher upper mass limit on our simulation results compared to the basic ONC model is investigated in
Fig.~\ref{fig:disc_mass_loss_vs_mass_bins__disc_size_fixed__basic_onc_model_50_msun_and_150_msun}. It shows the average relative disc-mass loss as a
function of the stellar mass for the standard ONC model with a stellar upper mass limit of 50\,\Msun\ (dark grey boxes), and the same model with an
upper mass limit of 150\,\Msun\ (light grey boxes). For masses below 10\,\Msun\ the two distributions are quantitatively in good agreement.

In the range of $\sim$10-50\,$\Msun$ the average disc-mass loss of the 50\,$\Msun$-limit model is significantly higher. This is to be expected because
stars in this mass range can act as additional strong gravitational foci in the presence of a 50\,$\Msun$ star, while their effect is largely reduced
if a 150\,$\Msun$ star is gravitationally dominating. The most massive stars in the range $80-150\,\Msun$ show, as expected, the largest average
relative disc-mass loss. That it is somewhat lower than that of the most massive star of the 50\,$\Msun$-limit model is in agreement with the stronger
gravitational attraction of their disc, leading on average to a reduced disc-mass loss per encounter. However, since the highest mass bin is only
populated by 9 stars, any further conclusions about the average relative disc-mass loss of the most massive stars would be highly speculative.

%

\section{Comparison of numerical results and analytical estimates}

\label{sec:analytical_results}

In this section the numerical results will be compared to analytical estimates. The treatment of encounters involves one important time scale, the
\emph{collision time} $t_{\mathrm{coll}}$ \citep[][Eq.~8-123]{1987gady.book.....B},
\begin{equation}
  \label{eq:tcoll}
  t_\mathrm{coll} = \left[ 16 \sqrt{\pi} \rho \sigma r_\star^2 \left(1 + \frac{Gm_\star}{2\sigma^2r_\star}\right) \right]^{-1} \,.
\end{equation}
Here the inverse of the collision time will be introduced as the encounter rate, $f_{\mathrm{enc}} \equiv t_{\mathrm{coll}}^{-1}$. Using the escape
velocity $v_{\star}$ from the stellar surface,
\begin{equation}
  \label{eq:escape_velocity_stellar_surface}
  v_{\star} = \sqrt{ \frac{ 2 G m_{\star} }{ r_{\star}} } \,,
\end{equation}
the encounter rate can be written as
\begin{equation}
  \label{eq:encounter_rate_relative}
  f_{\mathrm{enc}} 
  = 16 \sqrt{\pi} \rho \sigma r_\star^2 \left( 1 + \frac{ v_{\star}^2 }{ 4\sigma^2 } \right) \,,    
\end{equation}
where $G$ denotes the gravitational constant, $m_{\star}$ and $r_{\star}$ the stellar mass and radius, and $\rho$ and~$\sigma$ the density and
velocity dispersion of the star cluster. In the following, the stellar radius~$r_{\star}$ will be replaced by the ``typical interaction radius''
$r_{\mathrm{enc}}$, that means the radius at which the star is subject to a significant (but still frequent) perturbation that potentially can remove
some fraction of the disc-mass. Eq.~(\ref{eq:encounter_rate_relative}) will be evaluated for three different stellar masses, representing stellar mass
groups of low\mbox{-,} intermediate- and high-mass stars. Appropriate typical interaction radii have been taken from Table~3 of
\citet{2006ApJ...642.1140O}. The set of masses $m_{\star}$, radii $r_{\mathrm{enc}}$, and ``encounter escape speeds'' $v_{\mathrm{enc}}$ resulting
from Eq.~(\ref{eq:escape_velocity_stellar_surface}) is shown in Table~\ref{tab:encounter_rate_parameter_space}. The last column contains the
``gravitational focusing parameter'' $\gamma_{\mathrm{enc}}$, an approximation parameter defined as the power of ten best representing
$v_{\mathrm{enc}}^2$,
\begin{equation}
  \label{eq:gravitational_focusing_parameter}
  \gamma_{\mathrm{enc}} \equiv 10^{\displaystyle \, \lfloor \log{v_{\mathrm{enc}}^2} + 1/2 \rfloor} \,,
\end{equation}
where $\lfloor{x}\rfloor$ denotes the floor function\footnote{The floor function $\lfloor{x}\rfloor$ gives the largest integer less than or equal to
  $x$.} of $x$. For clarity, mass-dependent quantities, $x_{\mathrm{enc}}$, will be explicitly denoted as functions of mass,
$x_{\mathrm{enc}}(m_{\star})$, in the following.

\begin{table}
  \centering
  \begin{tabular}{l*4{c}}
    \hline
    mass groups        &  $m_{\star}$  &  $r_{\mathrm{enc}}$ &  $v_{\mathrm{enc}}$  & $\gamma_{\mathrm{enc}}$   \\
    &  [$\Msun$]   &  [AU]             &  [AU/yr]           &             \\
    \hline
    low mass           &  $0.1-1~~~~$     &  $10^{2-3~~}$        &  $\sim0.3$         &  $10^{-1}$  \\
    intermediate mass  &  $~~~1-10~~$      &  $10^{2-2.5}$      &  $\sim 2$          &  $1$        \\
    high mass          &  $~10-100$    &  $10^{2~~~~~}$           &  $\sim 6$          &  $10$       \\
    \hline
  \end{tabular}
  \caption{Typical parameters adopted for the calculation of the encounter rate of cluster stars. The first column denotes the three mass groups, the
    second column contains the adopted mass ranges, $m_{\star}$, while in the third typical interaction radii, $r_{\mathrm{enc}}$, are listed for each
    mass group. In the last two columns the resulting ``encounter escape speeds'', $v_{\mathrm{enc}}$, and the ``gravitational focusing parameter'', $\gamma_{\mathrm{enc}}$, are
    noted (see text for definitions of these quantities).}
  \label{tab:encounter_rate_parameter_space}
\end{table}

Because the model D2/S2 represents the standard ONC model, which has been intensively studied, the calculations will be normalised to this model. All
quantities related to this model will be thus denoted by a ``0'' as subscript. Adopting the initial velocity dispersion of the model D2/S2, $\sigma_0
\approx 2.3\,\kms \approx 0.5\,\mathrm{AU/yr}$, using
\begin{equation}
  4\sigma^2 =
  4\sigma_0^2 \left( \frac{ \sigma } { \sigma_0 } \right)^2  \approx
  \left( \frac{ \sigma } { \sigma_0 } \right)^2 \frac{\mathrm{AU}^2}{\mathrm{yr}^2} \,,
  \nonumber
\end{equation}
and the numbers given in Table~\ref{tab:encounter_rate_parameter_space}, Eq.~(\ref{eq:encounter_rate_relative}) can be simplified
to
\begin{equation}
  \label{eq:encounter_rate_relative_compact}
  f_{\mathrm{enc}} = 16 \sqrt{\pi} \rho \sigma r_{\mathrm{enc}}^2(m_{\star}) \left[ 1 + \gamma_{\mathrm{enc}}(m_{\star}) \left( \frac{ \sigma_0 } { \sigma } \right)^2 \right] \,.
\end{equation}
An even more compact representation is achieved by considering the scaling properties of the two families of models: as can be derived from
Eqs.~(\ref{eq:relation_number_density_size}) and (\ref{eq:relation_velocity_dispersion}), the scaling relations for the density-scaled models are
$\rho \propto N$ and $\sigma \propto \sqrt{N}$, while $\rho = \mathrm{const}$ and $\sigma = \mathrm{const}$ is found for the size-scaled models. Using
these relations, transforming $r_{\mathrm{enc}}(m_{\star})$ to $\gamma_{\mathrm{enc}}(m_{\star})$ via Eqs.~(\ref{eq:escape_velocity_stellar_surface})
and (\ref{eq:gravitational_focusing_parameter}), and normalising the encounter rate to the model D2/S2, $f_{\mathrm{enc}}^{\mathrm{norm}} \equiv
f_{\mathrm{enc}} / f_{\mathrm{enc,0}}$, one obtains
\begin{equation}
  \label{eq:encounter_rate_normalized}
  \begin{split}
    f_{\mathrm{enc}}^{\mathrm{norm}} = & \left( \frac{m_{\star}}{m_{\star,0}} \frac{\gamma_{\mathrm{enc}}(m_{\star,0})}{\gamma_{\mathrm{enc}}(m_{\star})} \right)^2 \times \\
    &
    \begin{cases}
      \left( \dfrac{\rho}{\rho_0} \right)^{3/2}
      \dfrac{1 + \left( \frac{\rho}{\rho_0} \right)^{-1} \gamma_{\mathrm{enc}}(m_{\star})} {1 + \gamma_{\mathrm{enc}}(m_{\star,0})}
      & \text{D-models} \,, \\[2ex]
      \phantom{\left( \dfrac{\rho}{\rho_0} \right)^{3/2}}
      \dfrac{1 + \gamma_{\mathrm{enc}}(m_{\star})} {1 + \gamma_{\mathrm{enc}}(m_{\star,0})}
      & \text{S-models} \,.
    \end{cases}
  \end{split}
\end{equation}

The derived relation for the normalised encounter rate predicts very different scaling relations for the two families of cluster models. Density
scaled-models are expected to show large variations of the number of encounters, with a strong dependency on the density for low-mass stars, i.e. when
$\gamma_{\mathrm{enc}}(m_{\star}) \ll 1$. In contrast, the encounter rate for the size-scaled models is expected to vary only for the different mass
groups, but not among different models. For a better overview of the scaling in terms of numbers, Table~\ref{tab:encounter_rate_normalized_tabulated}
lists approximated relative encounter rates $f_{\mathrm{enc}}^{\mathrm{norm}}$, normalised to the low-mass group of the model
D2/S2. Table~\ref{tab:encounter_rate_normalized_tabulated} demonstrates that the gravitational focusing parameter plays an important role for the
massive stars. The encounter rates increase dramatically by roughly one order of magnitude from the low- and intermediate-mass stars to the high-mass
stars for the model D2/S2. This finding agrees well with the number of encounters of the ONC model found earlier \citep[cf.][]{2006A&A...454..811P}.

\begin{table*}
  \centering
  \begin{tabular}{|l|*{6}{c}|c|}
    \hline
    family of models   & \multicolumn{6}{c|}{density-scaled}          & size-scaled \\
    cluster model      & D0    &  D1  &  D2  &  D3  &  D4  &  D5  &  S0-S5 \\
    \hline
    low mass           &  1/5  & 2/5  &  1   &  3   &  7   &  21  &  1    \\
    intermediate mass  &  1/2  & 1    &  2   &  4   &  9   &  23  &  2    \\
    high mass          &  5    & 7    &  10  &  15  &  25  &  44  &  10   \\
    \hline
  \end{tabular}
  \caption{Approximate relative encounters rates $f_{\mathrm{enc}}^{\mathrm{norm}}$ from Eq.~(\ref{eq:encounter_rate_normalized}) of the 
    density-scaled and size-scaled models, normalized to the low-mass group of the model D2/S2.}
  \label{tab:encounter_rate_normalized_tabulated}
\end{table*}

In summary, what one would expect from the numerical simulations of the density-scaled models is a steep increase of the encounter rate $\propto
\rho^{3/2}$ in case of the low-mass stars and a considerably shallower dependency $\propto \rho^{1/2}$ for the high-mass stars. For low densities,
corresponding to low particle numbers, one would expect that the high-mass stars dominate the encounter rate via gravitational focusing, favouring on
overall scaling $\propto \rho^{1/2}$. In contrast, the size-scaled models should produce very similar results in terms of encounter rate, independent
of the specific cluster model.

\begin{figure*}
  \centering
  \includegraphics[width=0.49\linewidth]{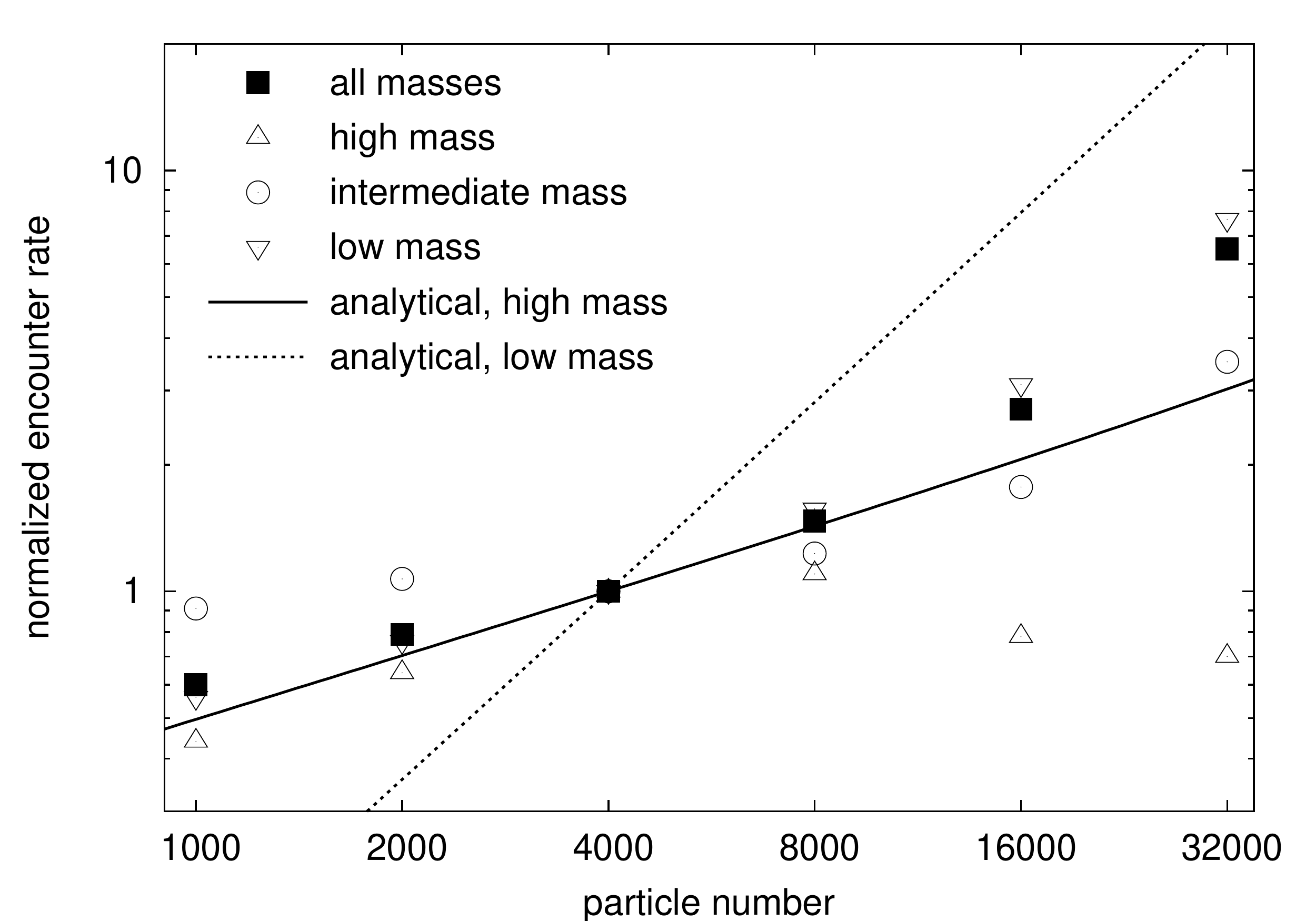}
  \includegraphics[width=0.49\linewidth]{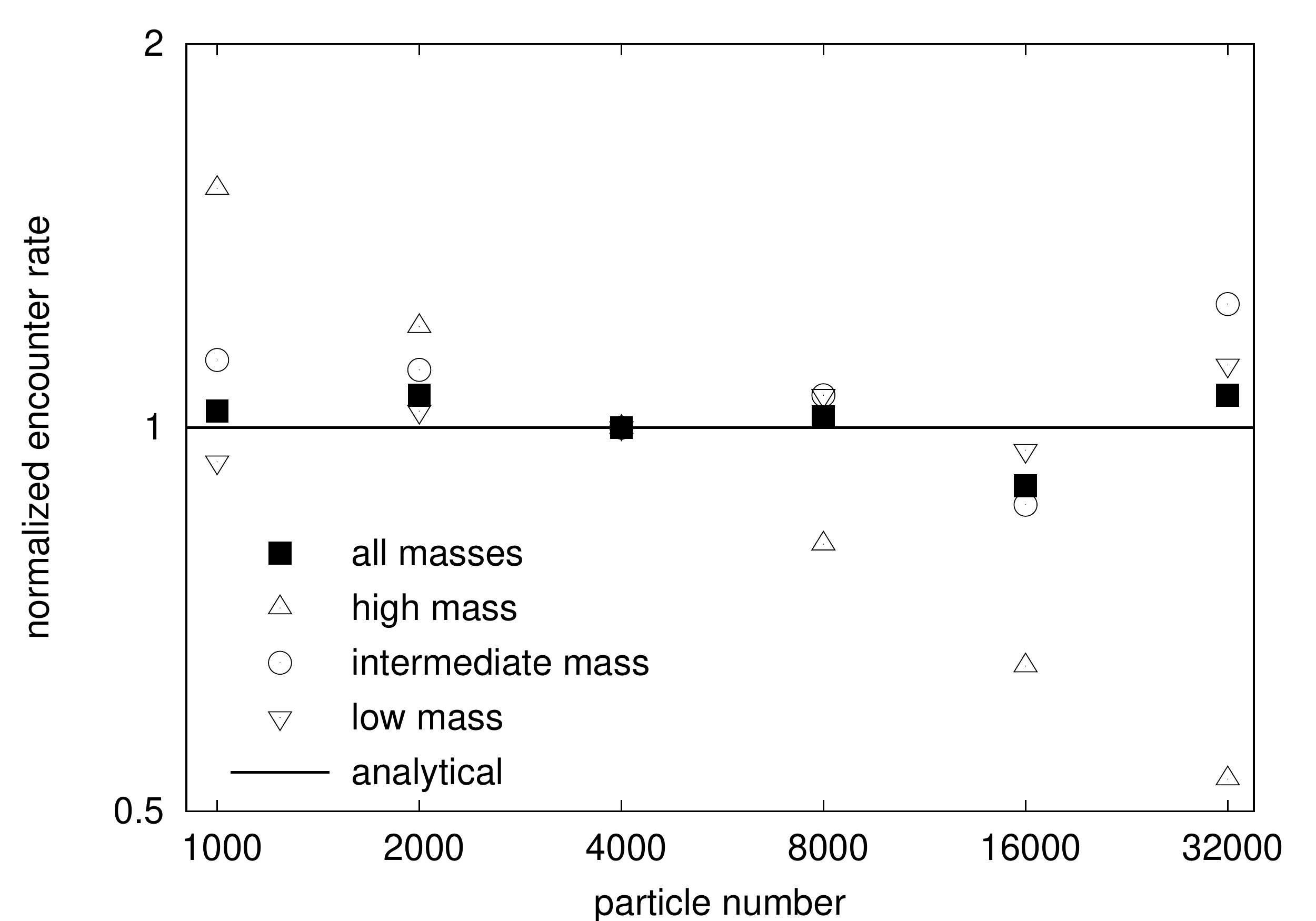}  
  \caption{Normalised encounter rate $f_{\mathrm{enc}}^{\mathrm{norm}}$ from the simulations in comparison with the analytical estimate given by
    Eq.~(\ref{eq:encounter_rate_normalized}). The filled squares represent all stars in the Trapezium Cluster region ($R=0.3\,\mathrm{pc}$), the other
    open symbols stand for predefined mass groups: high-mass, $m \ge 10\,\Msun$ (upward triangles), intermediate-mass, $10\,\Msun \ge m \ge 1\,\Msun$
    (circles), and low-mass stars, $m \le 1\,\Msun$ (downward triangles). The lines depict the analytical estimate of the encounter rate for high-mass
    (dashed line) and low-mass stars (solid line). The ranges of the mass groups have been chosen here different from those in previous figures to
    account for the mass regimes of the encounter rate presented in Table~\ref{tab:encounter_rate_parameter_space}. \emph{Left:} Density-scaled
    cluster models. \emph{Right:} Size-scaled cluster models (here the analytical estimates of the encounter rate for high-mass and low-mass stars are
    identical).}
  \label{fig:normalized_encounter_frequency}
\end{figure*}

These expectations are in good agreement with the results from the numerical simulations. We demonstrate this in
Fig.~\ref{fig:normalized_encounter_frequency} via the average encounter rate of stars of all masses and of the three mass groups, respectively,
normalised in each case to the model D2/S2. Fig.~\ref{fig:normalized_encounter_frequency}a shows that the encounter rates of the cluster models D0-D2
are scaling roughly as $N^{1/2}$. For higher particle numbers the distribution becomes more complex. Here the high-mass stars show a trend of
\emph{decreasing} encounter rate with particle number. This feature accounts for the decreasing importance of the high-mass stars as gravitational
foci (for the lower mass stars) and is a consequence of the decreasing ratio of the mass of the most massive star and the cluster mass. Accordingly,
the distribution of the encounter rate tends towards the analytical limit of $N^{3/2}$ for low-mass stars, representing the more frequent interaction
of low-mass stars with each other in the models D3-D5.

Fig.~\ref{fig:normalized_encounter_frequency}b shows that the size-scaled models are equivalent in their environmental effect on protoplanetary discs.
In the case of low- and intermediate mass stars the presented encounter rates, normalised to the model S2, are in good agreement with the analytical
estimate, which predicts a constant distribution as a function of particle number. In contrast, the normalised encounter rate of the high-mass stars
decreases with increasing particle number. This trend shows that the high-mass stars, similarly to the finding for the density-scaled models, become
less important as gravitational foci for the low-mass stars in clusters with larger stellar populations.

%

\section{Conclusion and Discussion}

\label{sec:conclusions}

The influence of different cluster environments on the encounter-induced disc-mass loss has been investigated by scaling the size, density and stellar
number of the basic dynamical model of the ONC. The findings can be summarized as follows:
\begin{enumerate}
\item The disc-mass loss increases with cluster density but remains rather unaffected by the size of the stellar population.
\item The density of the ONC itself marks a threshold:
  \begin{enumerate}
  \item in less dense and less massive clusters it is the massive stars that dominate the encounter-induced disc-mass loss via gravitational focusing of
    low-mass stars whereas
  \item in denser and more massive clusters the interactions of low-mass stars with equal-mass perturbers play the major role for the removal of disc
    mass.
  \end{enumerate}

\item In clusters four times sparser than the ONC the effect of encounters is still apparent.
\end{enumerate}

These findings from numerical simulations are well confirmed via observations. It is widely known that the disc frequency of even very young clusters
(e.g. NGC~2024) is significantly below 100\,\% \citep{2001ApJ...553L.153H,2005astro.ph.11083H}. This implies a very short time-scale for the disc
destruction of a part of the cluster population. In fact, stellar interactions are a potential candidate for a very rapid physical process and thus
the encounter-induced disc-mass loss is potentially a vital mechanism for disc destruction at the earliest stages of cluster evolution. Observational
evidence for the role of this mechanism has also been provided for the $\sim$1\,Myr old Orion Nebula Cluster \citep{2008A&A...488..191O}. Moreover,
observations of NGC~2024, ONC, and NGC~3603 confirm also the correlation of decreasing cluster disc fraction (CDF) with increasing cluster density and
even provide evidence for a critical density $\rho_{\mathrm{crit}} \approx \rho_{\mathrm{ONC}}$. A compilation of observational data shows that these
similar aged ($\sim$1\,Myr) clusters with peak densities of roughly $10^3\,\mathrm{pc}^{-3}$, $10^4\,\mathrm{pc}^{-3}$, and $10^5\,\mathrm{pc}^{-3}$,
respectively, show CDFs of $85\,\%$, $80\,\%$, and $40\,\%$ \citep{2000AJ....120.1396H,2000AJ....120.3162L,2004AJ....128..765S}. The fact that
NGC~2024 is sparser than the ONC but has a similar CDF is in good agreement with our simulations.

\begin{figure}
  \centering
  \includegraphics[width=1.0\linewidth]{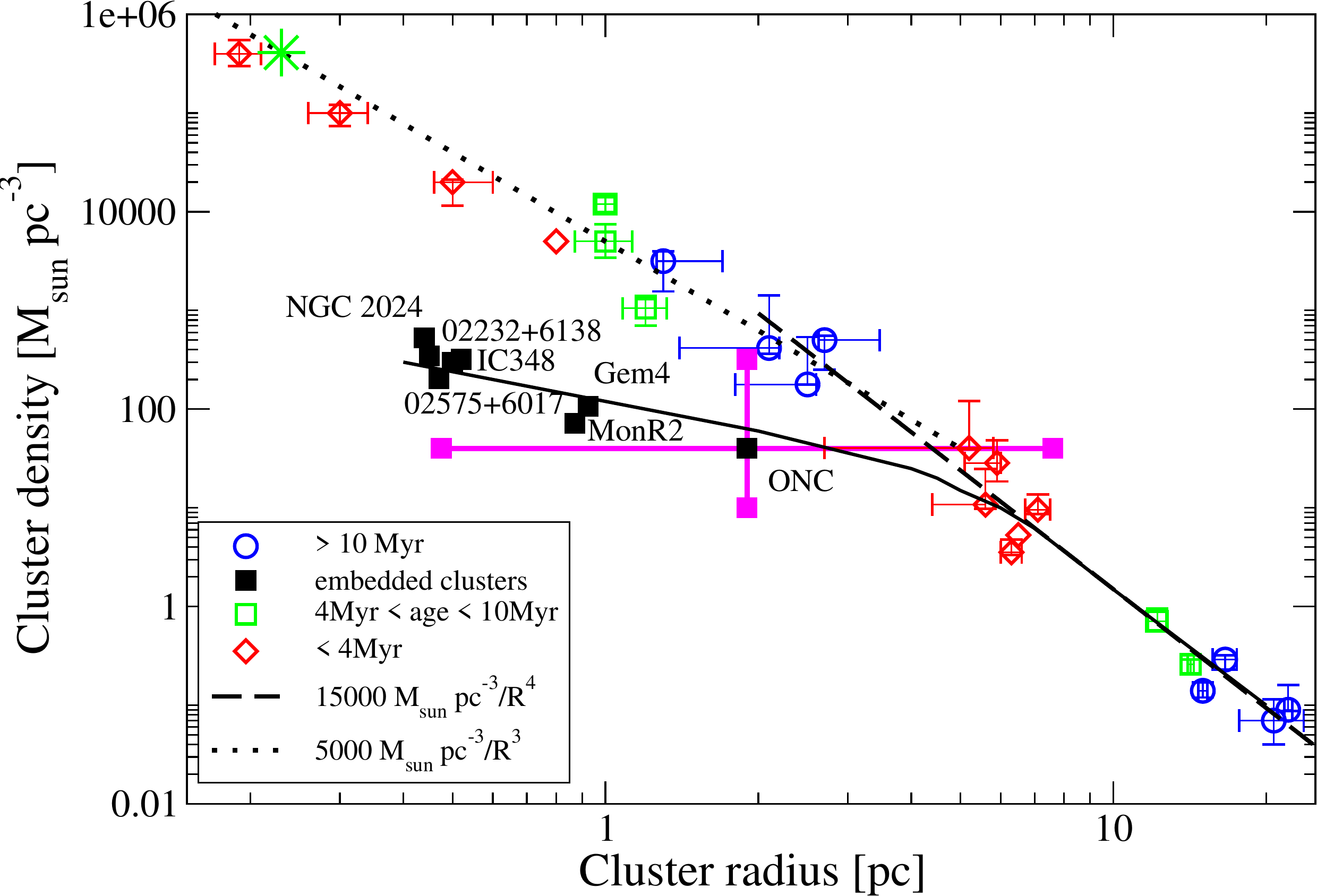}
  \caption{Cluster density as a function of cluster size for clusters more massive than $10^3\,\Msun$ and embedded clusters with more than 200
    observed members from \cite{2009A&A...498L..37P}. The parameter space covered by the simulations in this work is indicated by the large pink
    cross centered on the ONC.}
  \label{fig:cluster_sequences__simulated_parameter_space}
\end{figure}

Our results have several important implications for the general picture of star and cluster formation. Very recently, \cite{2009A&A...498L..37P} has
shown that there exist two cluster sequences evolving in time along pre-defined tracks in the density-radius plane, the ``leaky'' and the
``star-burst'' clusters. The simulations performed in the present investigation cover the parameter space of the ``leaky'' clusters in their embedded
stage (see Fig.~\ref{fig:cluster_sequences__simulated_parameter_space}). Comparison with our findings shows that at the earliest evolutionary stage
leaky clusters have densities above the critical density. Hence in leaky clusters star-disc systems are initially efficiently destroyed via encounters
that occur preferentially between low-mass stars. The ONC corresponds to an intermediate stage in the embedded phase of leaky clusters with the
transition towards a preferred disc-mass loss of high-mass stars via gravitational focusing of low-mass stars. At the final stage of the embedded
phase the encounter-induced disc-mass loss in leaky clusters ceases. Gravitational focusing by high-mass stars may still affect single objects yet at
this age is most probably exceeded by other disc-destructive processes like photoevaporation or planet formation. The effects in star-burst clusters
would be similar yet much more pronounced. In case of the Arches cluster one could expect stellar encounters to destroy the discs of most of the low-
and high-mass stars in several hundred thousand years. Combining our results with the finding of \cite{2003ARA&A..41...57L} that most stars are born
in clusters, it becomes evident that a significant fraction of all stars must have been affected by stellar encounters at the early evolutionary
stages of their hosting environment.

The application of our results to the dynamics of embedded clusters -- though obtained from simulations that do not contain gas -- is justified for
three reasons:
\begin{enumerate}
\item Rather than simulating the \emph{evolution} of leaky clusters (which would explicitly require the treatment of gas), we use our cluster models
  to map certain \emph{evolutionary stages} of the sequence of leaky clusters in terms of "dynamical snapshots". The dynamical effects of these
  numerical models are then used to estimate the effect of encounters in the observed clusters at their current dynamical state.
\item The effect of gas in an embedded cluster is to lower the frequency of close encounters (due to the smoother cluster potential), yet unless the
  gas mass is dominating cluster dynamics - as is not the case for the only partially embedded leaky clusters shown in
  Fig.~\ref{fig:cluster_sequences__simulated_parameter_space} (e.g. NGC~2024) - the effect is minor.
\item Gas expulsion causes the clusters to expand and thus their density to decrease much faster than in our simulations. Consequently, when mapping
  our results to the current dynamical state, we \emph{underestimate} the initial cluster density and thus the effect of encounters in the early
  evolutionary phases. Hence, our results are least accurate for older clusters, yet well applicable for the young evolutionary stages were encounters
  have the largest effect.
\end{enumerate}

The effect of the encounter-induced disc-mass loss in the early evolution of stellar systems has important implications for the formation and
evolution of planets. From our findings we conclude that the probability to find planets around intermediate mass stars is highest. While the high
initial densities of leaky clusters imply that planets around low-mass stars are expected to be less frequent, hardly any are expected in star-burst
clusters. Independent of the cluster environment, planet formation around high-mass stars seems to be completely hindered via interactions with other
cluster members.

%

\begin{acknowledgements}
  We thank the anonymous referee and the editor for careful reading and very useful comments and suggestions which improved this work. C. Olczak
  appreciates fruitful discussions with S. Portegies Zwart concerning the analytical estimates and scaling relations. We also thank R. Spurzem for
  providing the {\textsc{\mbox{nbody6\raise.2ex\hbox{\tiny{++}}}}} code for the cluster simulations. Simulations were partly performed at the
  J\"{u}lich Supercomputing Centre (JSC), Research Centre J\"{u}lich, Project HKU14. We are grateful for the excellent support by the JSC Dispatch
  team.
\end{acknowledgements}

%

\bibliographystyle{aa}
\bibliography{references}

%

\appendix

%
\section{Determination of boundaries of mass groups}

\label{app:star_cluster_models:mass_groups}

Boundaries of mass groups of low-, intermediate- and high-mass stars have been determined individually for different sizes of stellar populations on
the basis of the IMF of \citet[][see also Eq.~\ref{eq:kroupa_imf}]{2001MNRAS.322..231K}. The derivation involves the requirement for the three mass
ranges to be equidistant in logarithmic space, weighted by the slope of the IMF (of each mass range). The weighting accounts for the steepness of the
slope in the high-mass regime which would otherwise cause a very sparsely populated group of high-mass stars.

In the case of a lower mass cutoff at $m_0 = 0.08\,\Msun$, and an upper mass limit $m_3$, the IMF is characterised by just two different slopes,
$\alpha_1=1.3$ in the range $m_0 \le m < 0.50\,\Msun$, and $\alpha_2=2.3$ in the range $0.50\,\Msun \le m \le m_3$. Because the break in the slope of
the IMF at the critical mass $m_c^{\mathrm{br}} = 0.5\,\Msun$ does not necessarily coincide with one of the boundaries of the mass ranges, the cases
$m_1 < m_c^{\mathrm{br}}$ and $m_1 \ge m_c^{\mathrm{br}}$ have to be differentiated. Though from the theoretical point of view the same
differentiation would be required for the higher mass boundary $m_2$, this is not relevant for the stellar systems in the focus of the present
work. The four mass ranges, $m_k$, $k=0,..,3$, and the two slopes, $\alpha_k$, $k=1,2$, are then interrelated as follows:
\begin{equation}
  \nonumber
  \begin{split}
    (m_1& \ge m_c^{\mathrm{br}}) \wedge  (m_2 \ge m_c^{\mathrm{br}}): \\
    & \phantom{\equiv~} (\log{m_1} - \log{m_c^{\mathrm{br}}}) \alpha_2^{-1} + (\log{m_c^{\mathrm{br}}} - \log{m_0}) \alpha_1^{-1} \\
    & \equiv (\log{m_2} - \log{m_1})\alpha_2^{-1} \\
    & \equiv (\log{m_3} - \log{m_2})\alpha_2^{-1} \,,\\
    \\
    (m_1& < m_c^{\mathrm{br}}) \wedge (m_2 \ge m_c^{\mathrm{br}}): \\
    & \phantom{\equiv~} (\log{m_1} - \log{m_0}) \alpha_1^{-1} \\
    & \equiv (\log{m_2} - \log{m_c^{\mathrm{br}}}) \alpha_2^{-1} + (\log{m_c^{\mathrm{br}}} - \log{m_1}) \alpha_1^{-1} \\
    & \equiv (\log{m_3} - \log{m_2}) \alpha_2^{-1} \,.\\
  \end{split}
\end{equation}
\\
Solving these equations, and substituting
\begin{equation}
  \nonumber
  \begin{aligned}
    \alpha_{12}& \equiv \alpha_1 \alpha_2^{-1}, \\
    \alpha_{21}& \equiv \alpha_2 \alpha_1^{-1}, \\
    \Gamma_{12}& \equiv 1 - \alpha_{12}, \\
    \Gamma_{21}& \equiv 1 - \alpha_{21},
  \end{aligned}
\end{equation}
one obtains
\begin{equation}
  \label{eq:mass_groups_solution}
  \begin{split}
    (m_1& \ge m_c^{\mathrm{br}}) \wedge  (m_2 \ge m_c^{\mathrm{br}}): \\
    & \log{m_1} = \tfrac{1}{3} \left[   \log{m_3} + 2 \Gamma_{21} \log{m_c^{\mathrm{br}}} + 2 \alpha_{21} \log{m_0} \right] \,, \\
    & \log{m_2} = \tfrac{1}{3} \left[ 2 \log{m_3} +   \Gamma_{21} \log{m_c^{\mathrm{br}}} +   \alpha_{21} \log{m_0} \right] \,, \\
    \\
    (m_1& < m_c^{\mathrm{br}}) \wedge (m_2 \ge m_c^{\mathrm{br}}): \\
    & \log{m_1} = \tfrac{1}{3} \left[   \alpha_{12} \log{m_3} + \Gamma_{12} \log{m_c^{\mathrm{br}}} + 2             \log{m_0} \right] \,, \\
    & \log{m_2} = \tfrac{1}{3} \left[ 2             \log{m_3} + \Gamma_{21} \log{m_c^{\mathrm{br}}} +   \alpha_{21} \log{m_0} \right] \,. \\
  \end{split}
\end{equation}
\\
The choice of the appropriate solution is determined by the upper mass limit $m_3$. For this purpose the ``critical maximum mass''
$m_c^{\mathrm{max}}$,
\begin{equation}
  \nonumber
  m_c^{\mathrm{max}} = \log^{-1}{[ ( 1 + 2 \alpha_{12} ) \log{m_c^{\mathrm{br}}} - 2 \alpha_{12} \log{m_0} ]} \,,
\end{equation}
is estimated from Eq.~(\ref{eq:mass_groups_solution}) and $m_1 \equiv m_c^{\mathrm{br}}$. Consequently, the following relations hold:
\begin{equation}
  \nonumber
  \begin{aligned}
    m_3 <   m_c^{\mathrm{max}} \quad & \Longrightarrow \quad m_1 \ge m_c^{\mathrm{br}} \,, \\
    m_3 \ge m_c^{\mathrm{max}} \quad & \Longrightarrow \quad m_1 <   m_c^{\mathrm{br}} \,.
  \end{aligned}
\end{equation}
With the given values of the parameters $m_0$, $\alpha_1$, and $\alpha_2$ one finds
\begin{equation}
  \nonumber
  m_c^{\mathrm{max}} \approx 3.97\,\Msun \,.
\end{equation}

The derived mass boundaries, $m_k$, $k=0,..,3$, for each cluster of the families of models are presented in Table~\ref{tab:mass_groups}.
\begin{table*}
  \centering
  \begin{tabular}{l*6{c}}
    \hline
    \hline
    &  1000  &  2000  &  4000  &  8000  &  16000  & 32000 \\
    \hline
    $m_0 [\Msun]$  &
    $8.00 \cdot 10^{-2}$  &  $8.00 \cdot 10^{-2}$  &  $8.00 \cdot 10^{-2}$  &  $8.00 \cdot 10^{-2}$  &  $8.00 \cdot 10^{-2}$  &  $8.00 \cdot 10^{-2}$ \\
    $m_1 [\Msun]$  &
    $3.30 \cdot 10^{-1}$  &  $3.54 \cdot 10^{-1}$  &  $3.82 \cdot 10^{-1}$  &  $3.95 \cdot 10^{-1}$  &  $4.15 \cdot 10^{-1}$  &  $4.24 \cdot 10^{-1}$ \\
    $m_2 [\Msun]$  &
    $2.94$                &  $3.78$               &  $4.95$               &  $5.58$                &  $6.61$               &  $7.14$              \\
    $m_3 [\Msun]$  &
    $1.47 \cdot 10^{2}$   &  $1.47 \cdot 10^{2}$   &  $1.48 \cdot 10^{2}$   &  $1.48 \cdot 10^{2}$   &  $1.50 \cdot 10^{2}$  &  $1.50 \cdot 10^{2}$  \\
    \hline
  \end{tabular}
  \caption{Boundaries of the three mass groups of low-, intermediate-, and high-mass stars of cluster models with 1000,
    2000, 4000, 8000, 16000, and 32000 particles.}
  \label{tab:mass_groups}
\end{table*}



%

\end{document}